\documentclass[useAMS, usenatbib, fleqn]{mnras}
\pdfoutput=1
\usepackage{aecompl}
\usepackage{amsmath}
\usepackage{amssymb}
\usepackage{gensymb}
\usepackage{wasysym}
\usepackage{graphicx}
\usepackage{subfig}
\usepackage{array}
\usepackage{times}
\usepackage{url}
\usepackage{xcolor}
\usepackage{hyperref}
\hypersetup{colorlinks=true,citecolor=blue,linkcolor=blue,filecolor=blue,urlcolor=blue}

\def\mnras{MNRAS}
\def\apj{ApJ}
\def\apjl{ApJL}
\def\apjs{ApJS}

\def\aap{A\&A}
\def\aj{AJ}
\def\pasj{PASJ}
\def\nat{Nature}
\def\araa{ARA\&A}
\def\physrep{Phys.~Rep.}

\def\prd{Phys.~Rev.~D}

\def\ssr{Space~Sci.~Rev.}

\definecolor{orange}{rgb}{0.9,0.25,0}

\definecolor{add_green}{rgb}{0.13,.75,0.13}

\title[Splashback boundaries in hydrodynamic simulations]
    {The splashback boundary of haloes in hydrodynamic simulations}
\author[S.~O'Neil et al.]
    {\parbox{18cm}{Stephanie O'Neil$^{1}$ \thanks{E-mail: sloneil@mit.edu}, David J. Barnes$^{1}$, Mark Vogelsberger$^{1}$, Benedikt Diemer$^{2}$
     }\vspace{0.3cm}\\
     $^{1}$Department of Physics and Kavli Institute for Astrophysics and Space Research,
           Massachusetts Institute of Technology,
           Cambridge, MA 02139, USA \\
    $^{2}$Department of Astronomy, University of Maryland, College Park, MD 20742, USA
    }

\begin{document}

\date{Accepted 2021 April 20. Received 2021 April 10; in original form 2020 November 30 }

\pagerange{\pageref{firstpage}--\pageref{lastpage}}
\pubyear{2021}

\maketitle
\label{firstpage}

\begin{abstract}
The splashback radius, $R_{\rm sp}$, is a physically motivated halo boundary that separates infalling and collapsed matter of haloes.
We study $R_{\rm sp}$ in the hydrodynamic and dark matter only IllustrisTNG simulations.
The most commonly adopted signature of $R_{\rm sp}$ is the radius at which the radial density profiles are steepest.
Therefore, we explicitly optimise our density profile fit to the profile slope and find that this leads to a $\sim5\%$ larger radius compared to other optimisations.
We calculate $R_{\rm sp}$ for haloes with masses between $10^{13-15}{\rm M}_{\odot}$ as a function of halo mass, accretion rate and redshift.
$R_{\rm sp}$ decreases with mass and with redshift for haloes of similar $M_{\rm200m}$ in agreement with previous work.  We also find that $R_{\rm sp}/R_{\rm200m}$ decreases with halo accretion rate.
We apply our analysis to dark matter, gas and satellite galaxies associated with haloes to investigate the observational potential of $R_{\rm sp}$.
The radius of steepest slope in gas profiles is consistently smaller than the value calculated from dark matter profiles.
The steepest slope in galaxy profiles, which are often used in observations, tends to agree with dark matter profiles but is lower for less massive haloes.
We compare $R_{\rm sp}$ in hydrodynamic and \textit{N}-body dark matter only simulations and do not find a significant difference caused by the addition of baryonic physics.
Thus, results from dark matter only simulations should be applicable to realistic haloes.

\end{abstract}


\section{Introduction}

In our current understanding of structure formation, small matter perturbations in our early universe grow over cosmic time via gravity to form the structures we observe today \citep[e.g.][]{Rees1977,White1978, Blumenthal1984}.
These structures come in a variety of geometries, e.g. haloes, filaments and sheets \citep[e.g.][]{Bond1996b}.
Because these structures are woven together in a ``cosmic web" and not distinct objects, any study that relies on knowing the size or extent of an object implicitly relies on a choice establishing the meaning of size.

Commonly, a halo is defined as a sphere within which the average density is a constant multiple of some reference density.
Its mass is then the enclosed mass and its radius is that of the sphere.
Typically, the reference density is either the critical or mean density of the universe, $\rho_{\rm c}$ or $\rho_{\rm m}$ respectively.
The critical density of the universe is given by the total energy density corresponding to the density required for a flat universe:
\begin{equation}
    \indent\rho_{\rm c} = \frac{3H^2(t)}{8\pi G}\:,
\end{equation}
where $H$ is the time-dependent Hubble factor and $G$ is the gravitational constant.
The mean density of the universe refers to only the matter component of the energy budget, $\Omega_{\rm m}\rho_{\rm c}$ with $\Omega_{\rm m}$ corresponding to the matter fraction of the universe.
The enclosed density is typically defined as $200$, $500$ or $2500$ times the reference density.

A value of $200$ times the critical density is close to the value predicted by the spherical top-hat collapse model \citep{Gunn1972, Gunn1977, Peebles1980}, in which a spherical constant overdensity detaches from the Hubble flow to collapse into a halo.
In this scenario, the amount of mass required for the overdentiy is roughly $200$ times the critical density \citep[e.g.][]{Bryan1998}.

However, there are a number of problems that arise from using the spherical overdensity definition of halo size in realistic situations.
Defining a halo based on a fixed ratio to a reference density like $R_{\rm200m}$ causes a halo to experience pseudoevolution, i.e. change in size due to the evolution of the reference density \citep[e.g.][]{Diemer2013}.
As the universe expands, $\rho_{\rm m}$ decreases, so the radius of a halo must be adjusted to keep the enclosed density at a constant multiple of $\rho_{\rm m}$.
This occurs even if the halo itself is unchanging, so the spherical overdensity definition does not necessarily reflect the dynamical nature of the halo \citep[e.g.][]{Cuesta2008}.

Additionally, all parts of a halo do not collapse simultaneously as assumed in the spherical top-hat model.
\citet{Bertschinger1985} modelled the infall of shells around an overdensity and showed that caustics form for collisionless fluids.
Infalling shells tend to pile up near the apocentre of their orbits where their velocities are smaller.
This results in a steepening of the density profile at the boundary of the halo.
This motivates an alternative way to identify the boundary of a halo: that which encloses the first orbits of infalling particles.
This is referred to as the ``splashback radius" \citep[e.g.][]{Diemer2014, Adhikari2014, More2015} and is what we attempt to measure in this work.

The location of the splashback radius as determined by the caustic depends on the dynamics of the halo.
\citet{Adhikari2014} motivated the relationship between the accretion rate of a halo and the orbital radius of infalling particles.
Collapsing matter adds mass to the halo; therefore, mass collapsing at a later epoch enters a deeper potential which shrinks its orbit.
Increasing the mass of the halo also increases the enclosed overdensity, which contributes to the decrease in its position relative to the spherical overdensitiy definition.

In idealised spherical environments, the caustic marking the splashback radius occurs in the same location as the first apocentre of orbits.
In practice, the caustic and first apocentre are not the same, although they are correlated.
In phase-space, infalling material forms streams that are visible as caustics in the density profile \citep{Vogelsberger2009}.
\citet{Diemer2017b} traced the trajectories of particles in \textit{N}-body simulations and found that the averaged first apocentre occurs at a similar radius as the maximal rate of change in the density profile, i.e. the steepest slope.
Although the two features are similar, the steepest slope is not the strict definition of the splashback radius.
The trajectories of particles, however, cannot be observed, so the point of steepest slope in a halo's density profile is useful as an observable proxy for $R_{\rm sp}$.
\citet{More2015} used the steepest slope in spherically averaged density profiles to relate the splashback radius to the accretion rate and found a decrease in $R_{\rm sp}/R_{\rm200m}$ with accretion rate similar to what would be expected of the splashback radius calculated from particle trajectories.

To identify the splashback radius from a halo's density profile, it is necessary to have a good description of the density profile.
\citet{Navarro1996} fit a density profile across haloes with sizes ranging from dwarf galaxies to galaxy clusters in \textit{N}-body simulations and found that they all had a similar spherically averaged density profile.
More recently, the \citet{ Einasto1965, Einasto1969} profile has  been  used to describe  dark matter halo density profiles \citep[e.g.][]{Merritt2006, Graham2006, Gao2008, Stadel2009, Navarro2010, Ludlow2011}.
The caustic expected in the density profile near the first apocentre of orbits, however, is expected at a radius larger than the region of haloes that is generally well described by these profiles.

Recently, there has been an effort to understand the behaviour of dark matter haloes at larger radii where the steepening in the density profile would be apparent.
\citet{Diemer2014} studied the density profiles out to radii of $\sim9\:\rm{R}_{\rm vir}$ and detected the caustic.
They proposed a density profile that combines an Einasto profile with a term describing the gradual approach to the mean density of the universe that captures the drop in density.
The proposed density profile has been used to find the steepest point in spherically averaged dark matter, stellar and subhalo profiles in simulations \citep[e.g.][]{More2015, Fong2018, Xhakaj2020, Deason2020b} and in observational work using galaxies as traces of dark matter density \citep{More2016, Baxter2017, Chang2018, Shin2019, Murata2020}.

If the splashback radius can be robustly identified, it may provide a more physical boundary to characterise the size of a given halo.
However, while the majority of a halo's mass is dark matter, it is not the most easily observed component of a halo and we must link our results to observable quantities.
For spherical overdensity definitions, there exist well known relations between observable properties and halo mass \citep{Kaiser1986, Allen2011, Kravtsov2012, Biffi2014, Baxter2015, Barnes2017a, Barnes2017b, Bocquet2019, Barnes2020a}.
Baryons have additional physical processes associated with them \citep[e.g.][]{Pike2014,McCarthy2017} that influence the relationship between observable quantities and collisionless orbital dynamics.
With the development of large hydrodynamic simulations \citep[e.g.][]{Schaye2015,Vogelsberger2014a, Springel2018, Vogelsberger2020}, it is now possible to study these processes in a cosmological context.

Similarly to dark matter, subhaloes are also largely collisionless and galaxies have been used to trace the density profile of clusters \citep[e.g.][]{More2016,Baxter2017,Zuercher2019}.
These measurements extend sufficiently far to detect the steepening in the density profile and identify the splashback radius.
However, the formation of a galaxy deepens the potential of a subhalo, impacting its resistence to tidal stripping and dynamical friction.
Additionally, the processes of galaxy formation alter a galaxy's colour during its infall into a larger object.
Combined with issues surrounding membership determinaton \citep[e.g.][]{Koester2007,Rozo2007,Rykoff2014,Klein2019}, there are potential effects not captured by dark matter only simulations that may impact the determination of the splashback radius.
For example, \citet{Deason2020a} studied the splashback radius in Local Group simulations by studying both the dark matter and subhaloes.
They found that the subhaloes tend to produce a caustic in the density profile at a smaller radius than the dark matter for Local Group-\textit{like} haloes.
\citet{Xhakaj2020} also found that the subhalo profiles yield smaller values for the splashback radius for haloes with $M\sim10^{14}\:\rm{M}_{\odot}$ and subhaloes with $M\sim10^{12}\:\rm{M}_{\odot}$.
\citet{Deason2020b}, however, found that the stellar component of haloes, including star particles not bound within subhaloes, resulted in density profiles similar to those of the dark matter component in cluster mass haloes.

In addition to stars and galaxies, gas is also a major observable component of halos that can be detected through X-ray and the Sunyaev-Zel'dovich effect \citep[e.g.][]{Reiprich2013}.
Gas, however, is collisional and thus not expected to be governed by the same orbital dynamics as dark matter.
Instead of orbital dynamics, the gaseous boundary of a halo is expected to be determined by the shock jump conditions.
However, it is not unreasonable to expect that, at some level, the gas profile correlates with the dark matter profile even if the two components are not exactly aligned \citep[e.g][]{Lau2015}.

If the splashback radius can be measured reliably, it is a potential observable signature of otherwise hidden physics such as dynamical friction \citep{Adhikari2016}, galaxy evolution in clusters \citep{Adhikari2020}, the nature of dark matter \citep{Banerjee2020}, and deviations from General Relativity \citep{Adhikari2018, Contigiani2019}.

While in reality, haloes are not perfectly spherical \citep[e.g.][]{Fillmore1984, Sheth2001} and it is possible to define a non-spherical ``splashback shell" \citep[e.g.][]{Mansfield2017}, we use the spherically averaged profiles since they are more easily related to observations and provide a defined ``size" of a halo based on a single radius.

The goal of this paper is to explore the impact of baryons, and their associated processes, on the splashback radius using the IllustrisTNG simulations.
We use the point of steepest slope as a proxy for the splashback radius and refer to it as $R_{\rm sp}$ throughout the paper.
Using both the full physics and the dark matter only runs, we can compare the determination of the splashback radius from different physical components, i.e. total matter, dark matter, gas and galaxies.
First, we will establish a robust method for identifying the splashback radius and the inherent bias associated with a chosen method.
Then we will examine the location of the splashback radius in haloes as a function of halo mass and accretion rate for the different components.
We will compare results from hydrodynamic simulations and \textit{N}-body simulations and explore if the astrophysical processes associated with baryons impact the results obtained in \textit{N}-body simulations.
We do not consider observational constraints in this paper, and leave this to future work.

The rest of this paper is structured as follows.
In Section \ref{sec:methods}, we briefly describe the IllustrisTNG simulations, our halo selection, and our method for identifying the splashback radius.
In Section \ref{sec:results}, we compare the splashback radius in the density profiles of dark matter, gas, and galaxies as a function of halo mass, halo accretion rate, and redshift.
We summarise our conclusions in Section \ref{sec:conclusions}.

\section{Methods}
\label{sec:methods}
In this paper, we explore the location of the splashback feature, $R_{\rm sp}$ as a function of halo mass, halo accretion rate, and redshift.
In this Section, we briefly introduce the IllustrisTNG simulations, outline our halo and galaxy selection methods, and define our procedure for measuring the splashback feature.

\subsection{Simulations}
\label{sec:methods_simulations}
We analyse the largest simulation volume of the IllustrisTNG Project, TNG300, as described in \citet{Nelson2018, Marinacci2018, Springel2018, Naiman2018, Pillepich2018b}.
The IllustrisTNG Project is a suite of cosmological magnetohydrodynamic simulations performed using the moving-mesh code \textsc{Arepo} \citep{Springel2010,Weinberger2020} and an updated version of the Illustris galaxy formation model \citep{Vogelsberger2013,Torrey2014}.  This produces a range of galaxy types and realistic clusters \citep[e.g.][]{Vogelsberger2014a, Vogelsberger2018, Barnes2018a, Genal2018,  Donnari2020}.

All runs use cosmological parameters consistent with \cite{PlanckCollaborationXIII2016}: $\Omega_{\rm m} = \Omega_{\rm{dm}}+\Omega_{\rm{b}} = 0.3089,\ \Omega_{\rm{b}} = 0.0486,\ \Omega_{\Lambda} = 0.6911,\ \sigma_8=0.8159,\ n_s=0.9667$, and Hubble constant $H_0=100h\,\rm{km} \,\rm{s}^{-1}\,\rm{Mpc}^{-1}$ where $h=0.6774$.
TNG300 is a periodic cube with a side length of $302\,\rm{Mpc}$ and was run at three resolution levels.
We focus on the highest resolution run, TNG300-1 in the main body of this work.
For details of the lower resolutions runs, TNG300-2 and TNG300-3, as well as convergence tests, we refer the reader to Appendix \ref{apx:convergence}.

TNG300-1 has $2\times2500^3$ cells/particles, with a target gas cell mass of $1.1\times10^7\,\rm{M}_{\odot}$ and a dark matter particle mass of $5.9\times10^7\,\rm{M}_{\odot}$.
The gravitational softening length of the dark matter particles is $1.5\,\rm{kpc}$ in physical (comoving) units for $z\leq1$ $(z>1)$.
The gas cells utilise an adaptive comoving softening that reaches a minimum of $0.37\,\mathrm{kpc}$.
The corresponding dark matter-only simulation, TNG300-1-DM, removes the gas cells but has the same number of dark matter particles as its hydrodynamic counterpart.
The dark matter only run has a particle mass of $4.7\times10^{7}\rm{M}_{\odot}$.

The galaxy formation model is an evolution of the original Illustris project \citep{Vogelsberger2014b}.
This includes a re-calibrated supernova wind model \citep{Pillepich2018a}, a new radio mode active galactic nuclei (AGN) feedback scheme \citep{Weinberger2017}, and further refinements that improve the convergence of the numerical scheme \citep{Pakmor2016}.
Gas cells that are radiatively cooling, with metal line cooling contributions, form stars stochastically via the \citet{Springel2003} two-phase effective equation of state model.
These stars evolve and return mass, metals and energy to their surroundings via winds from asymptotic giant branch stars and supernovae.
Following \citet{DiMatteo2005}, $1.2\times10^6\rm{M}_{\odot}$ black holes are seeded in haloes that reach a mass of $7.4\times10^{10}\rm{M}_{\odot}$.
Black holes then accrete gas following an Eddington limited Bondi prescription and can grow through mergers with other black holes.
Feedback is injected into the surrounding medium in either the quasar mode or kinetic mode depending on the accretion rate, see \citet{Weinberger2017} for further details.

\subsection{Halo properties}
\label{sec:methods_properties}
We stack haloes by either mass or accretion rates to calculate the density profiles of the stacked set.
When plotting quantities as a function of mass or accretion rate, we use the median value within a bin for the set of haloes.

We can define the size of a halo to be the radius $R_{\rm200m}$ of a sphere with enclosed density 200 times the mean density of the universe $\rho_{\rm m}$ such that
\begin{equation}
 \indent R_{\rm200m} = \left(\frac{3M_{\rm200m}}{4\pi200\rho_{\rm m}}\right)^{1/3}
\end{equation}
and the halo mass is then $M_{\rm200m}$, the mass enclosed within $R_{\rm200m}$.  When referring to a halo's mass throughout the paper, we use $M_{\rm200m}$.

\begin{table*}
\begin{tabular}{c|ccccccccccc}
       Simulation & \multicolumn{11}{c}{$z$} \\
      & 0.0 & 0.1 & 0.2 & 0.3 & 0.4 & 0.5 & 0.6 & 0.7 & 0.8 & 0.9 & 1.0 \\
     \hline
     TNG300-1 & 1401 & 1447 & 1496 & 1558 & 1578 & 1568 & 1568 & 1546 & 1506 & 1459 & 1393 \\
     TNG300-1-DM & 1529 & 1568 & 1634 & 1692 & 1725 & 1720 & 1693 & 1682 & 1647 & 1607 & 1547 \\
\end{tabular}
\caption{The number of haloes included in our sample for each redshift in the full physics (TNG300-1) and dark matter only (TNG300-1-DM) high resolution runs.  This is the number of Friends-of-Friends groups with $M_{\rm200m}>10^{13}\rm{M}_{\odot}$.  Haloes within $10\times R_{\rm200m}$ of a larger halo have been removed from the sample.  See Table \ref{table:res_tot_sample_size} for the sample sizes in the medium and low resolution runs.}
\label{table:tot_sample_size}
\end{table*}

\subsubsection{Accretion rates}
\label{sec:methods_accretion}
Past work on the splashback radius has found a strong correlation between $R_{\rm sp}/R_{200\rm{m}}$ and the mass accretion rate of haloes \citep[e.g.][]{Diemer2014, Adhikari2014, More2015, Mansfield2017, Diemer2017b}.
We compute the accretion rate $\Gamma$ of a halo using
\begin{equation}
    \indent\Gamma = \frac{\Delta\log_{10} M}{\Delta\log_{10} a}\:,
    \label{eq:acc_rate}
\end{equation}
where $M$ is the mass of the halo and $a$ is the scale factor.
For consistency with our mass definition, we use $M=M_{\rm200m}$.

The value of the accretion rate depends on the time interval used to compute the change in mass and scale factor.
We follow \citet{Diemer2017b} and measure the accretion rate over one dynamical time, which corresponds to approximately a halo crossing time.
We define the dynamical time $t_{\rm dyn}$ as a function of redshift $z$ following \citet{Diemer2017a}:
\begin{equation}
    \indent t_{\rm dyn}(z) = 2^{3/2}t_{\rm H}(z)\left(\frac{\rho_{\Delta}(z)}{\rho_{\rm c}(z)}\right)^{-1/2}\:,
    \label{eq:t_dyn}
\end{equation}
where $t_{\rm H}$ is the Hubble time and $\rho_{\Delta}$ is the average enclosed density. 
Since we use $M_{\rm200m}$ as our mass,  $\rho_{\Delta}/\rho_{\rm c}=200\Omega_{\rm m}$ by definition.  Then $t_{\rm dyn}$ simplifies to:
\begin{equation}
     \indent t_{\rm dyn} = \left(\frac{2^{3/2}}{\sqrt{200\Omega_{\rm m}}}\right)t_{\rm H} = \frac{t_{\rm H}}{5\sqrt{\Omega_{\rm m}}}\:.
     \label{eq:t_dyn_reduced}
 \end{equation}
This time interval is therefore the same for all haloes at a given redshift.

To compute the accretion rate of a halo at a given redshift, we first calculate the dynamical time.
We then select the snapshot earlier in the simulation with a time difference closest to the calculated dynamical time.
We use the SubLink merger trees \citep{Rodriguez-Gomez2015} to identify the main progenitor of the halo.
The SubLink algorithm tracks subhaloes through the simulation by linking subhaloes in consecutive snapshots that share the most bound particles.
The haloes are linked by associating each halo with its most bound subhalo.
We then calculate the change in mass between the snapshots as well as the change in scale factor.

\begin{figure*}
    \centering
    \includegraphics[width=\linewidth]{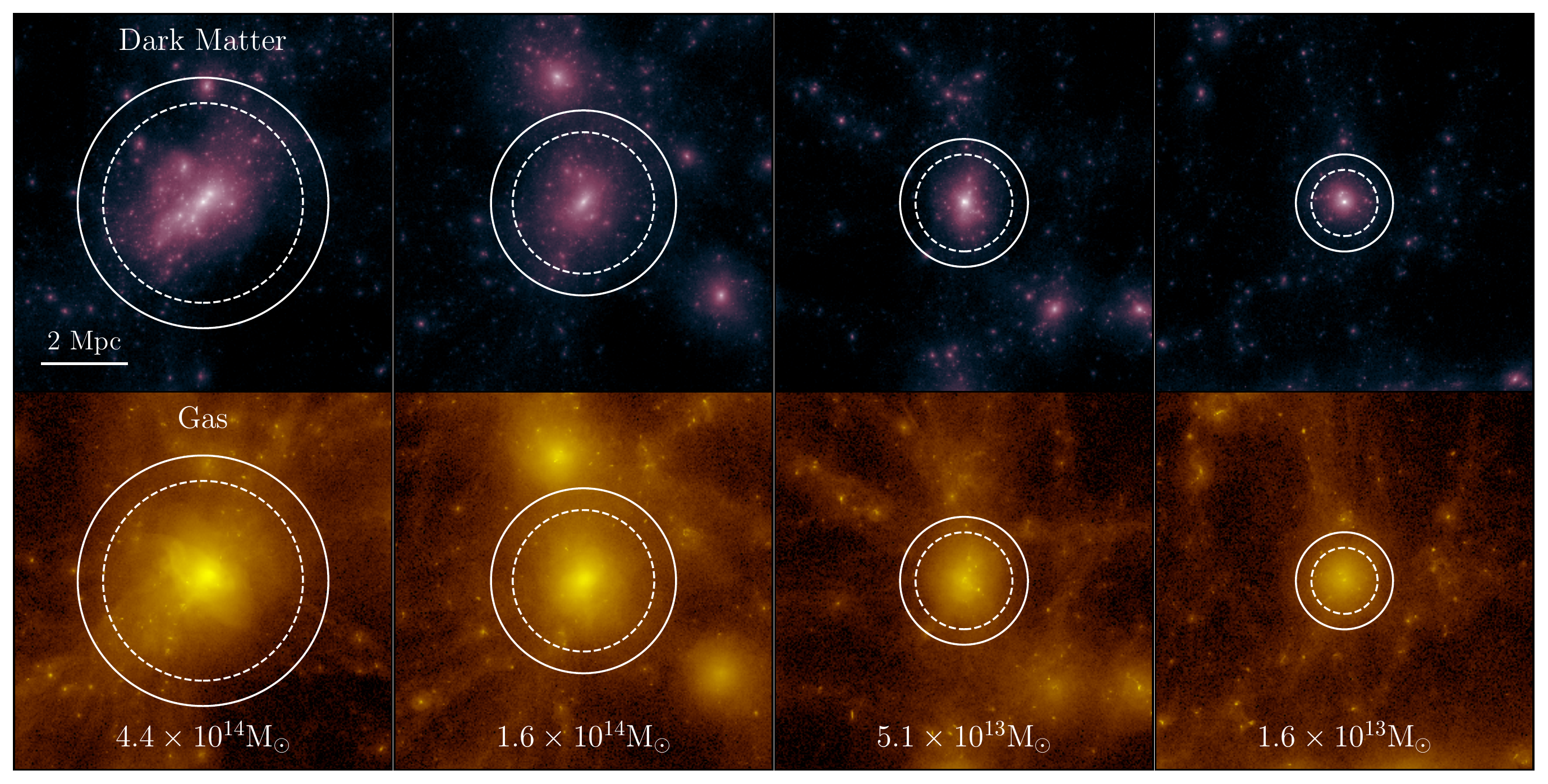}
    \caption{Projections for the dark matter (top) and gas (bottom) density for representative haloes in our samples.  We split the haloes into four logarithmically spaced mass ranges and show the median mass halo in each mass range.  $R_{\rm200m}$ is shown with the dashed line.  Our method for identifying the splashback feature is not robust for single haloes, so we calculate $R_{\rm sp}$ for the stacked dark matter profile of haloes in each mass range and show this with the solid line.  These demonstrate the difference between $R_{\rm200m}$ and $R_{\rm sp}$ shown with representative haloes.}
    \label{fig:halo_splashback}
    \vspace{.2cm}
\end{figure*}

\subsection{Halo and galaxy samples}
\label{sec:methods_samples}

We select samples of haloes from simulation snapshots over the redshift range $0\leq z\leq1$.
Gravitationally bound structures are identified within the simulation via the \textsc{Subfind} algorithm \citep{Springel2001,Dolag2009}.
These are associated with each other and additional particles using a Friends-of-Friends (FoF) algorithm \citep{Davis1985}.  This identifies structures by linking particles with neighbouring particles that lie within a linking length of $b=0.2$.
The most massive gravitationally bound object in a FoF group is identified as the main halo and others are labelled as subhaloes.
The centre of the halo is defined as the position of the most bound particle within the main halo, as determined by \textsc{Subfind}.
For the purposes of this paper, we define a galaxy as a subhalo with $M_{\rm200m}>10^9\rm{M}_{\odot}$ and a non-zero stellar mass.

We select an initial sample of haloes at a given redshift by taking all haloes with a total mass $M_{\rm200m}>10^{13}\rm{M}_{\odot}$.
Given our interest in effects occurring beyond the outskirts of the halo, we remove haloes from our sample that fall within $10R_{\rm200m}$ of a more massive object.
This tends to exclude haloes in more dense environments and with lower accretion rates due to stronger tidal forces, but it also ensures that the haloes in our sample are not being disrupted by larger objects.
With this selection criteria, we obtain 1401 haloes at $z=0$ and 1393 haloes at $z=1$ from the TNG300-1 simulation.
For the number of haloes at redshifts between 0 and 1, see Table \ref{table:tot_sample_size}.

\begin{table*}
\begin{tabular}{c|c|cccc|cccccccc}
     Simulation & $z$ & \multicolumn{4}{c|}{$\log_{10}\left(M_{\rm halo}\,/\,{{\rm M}_{\odot}}\right)$} & \multicolumn{8}{c}{$\Gamma$} \\
     & & \!{13.0-13.5}\! & \!{13.5-14.0}\! &  \!{14.0-14.5}\! & \!{14.5-15.0}\! & \!0.0-0.5\! & \!0.5-1.0\! & \!1.0-1.5\! & \!1.5-2.0\! & \!2.0-2.5\! & \!2.5-3.0\! & \!3.0-4.0\! & \!4.0-5.0\! \\
     \hline
     &  $0.0$  & 770 & 394 & 185 & 47 & 
                  109 & 361 & 316 & 237 & 154 & 90  & 87  & 25 \\
     TNG300-1 &  $0.5$  & 991 & 444 & 119 & 13 & 
                  83  & 197 & 267 & 222 & 182 & 176 & 228  & 127 \\
     &  $1.0$  & 1005 & 336 & 50  & 2  & 
                  39   & 102 & 152 & 145 & 168 & 157 & 243 & 198 \\
    \hline
    &   $0.0$  & 845 & 445 & 186 & 48 & 
                  140 & 447 & 343 & 235 & 140 & 111 & 81  & 21\\
    \! TNG300-1-DM\! &   $0.5$  & 1110 & 478 & 119 & 12  & 
                  79   & 258 & 301 & 250 & 207 & 188 & 251  & 119 \\
    &   $1.0$  & 1136 & 360 & 49  & 2   & 
                  44   & 109 & 207 & 180 & 184 & 173 & 286 & 196 \\
\end{tabular}
\caption{The number of haloes found in the high resolution hydrodynamic (TNG300-1) and \textit{N}-body (TNG300-1-DM) simulations for mass and accretion rate thresholds for select redshifts. See Table \ref{table:res_sample_size} for the sample sizes in the medium and low resolution runs. }
\label{table:sample_size}
\end{table*}

\subsection{Identifying the splashback feature}
\label{sec:methods_Rsp}
The splashback radius has been identified in a variety of ways.
\citet{Diemer2014} and \citet{More2015, More2016} used the steepest point in a spherically averaged density profile as the feature identifying  the splashback radius.
\citet{Diemer2017b} and \citet{Diemer2020a} instead traced the trajectories of individual dark matter particles as they fall in to haloes, selecting a radius that encloses the first apocentre of a percentage of their orbits.
The splashback radius is then defined as the smoothed average of the apocentre radii of individual particles.
\citet{Mansfield2017} sampled the density field around individual haloes to identify a ``shell'' around a halo based on the steepest point in the density profile along various lines of sight.
This creates a non-spherical boundary instead of defining a single radius.
In idealised spherical collapse models, the apocentre of all particles occurs at the same radius, creating a caustic in the density profile \citep{Huss1999, Adhikari2014}.  This manifests itself as a minimum of the density profile, although there is not a one-to-one match in practice.

In this work, we follow the first method and compute the location of the splashback radius as the steepest rate of change of stacked, spherically averaged density profiles within a given mass or accretion rate range.
The steepest slope of a density profile does not necessarily indicate the presence of the splashback feature; any halo will have a steepest point in its density profile.
The splashback feature is associated with the orbit of matter in a halo and the logarithmic slope decreasing significantly, typically below a value of -3.
However, we do not trace particle dynamics in this work to test the association between the steepest slope and particle orbits.

We also examine the density profiles of gas in haloes.
Though the gas density follows a similar shape to the dark matter profiles (see Figure \ref{fig:stacked_density_examples}), there are important differences.
Gas, as a collisional fluid, is not subject to the same orbital dynamics as dark matter and stars.
Given that there is an identifiable characteristic minimum in the gas profile similar to the dark matter profiles, however, we find it useful to investigate the location of steepest slope in the gas profile in our study of halo boundaries.
The method to identify this feature in the gas profiles follows the method to identify the feature in dark matter haloes, and it has the potential to provide a valuable basis for comparison to observations in the future.
We refer to the point of steepest slope as $R_{\rm sp}$ throughout the paper, keeping in mind that the underlying physics differs for the gas.

To get an idea of the difference between $R_{\rm sp}$ and $R_{\rm200m}$, we show the two radii relative to the dark matter and gas densities of four representative haloes in Figure \ref{fig:halo_splashback}.
We calculate $R_{\rm sp}$ from a stacked density profile of similarly sized haloes, with the haloes shown in the figure corresponding to the median mass of the stacked haloes.
$R_{\rm sp}$ (solid circle) tends to be larger than $R_{\rm 200m}$ (dashed circle) by $\sim20-50\%$.

\subsubsection{The density profile of haloes}
\label{sec:methods_density}
We compute the profile of a halo about its centre as defined by the halo finder descibed in Section \ref{sec:methods_samples}.
We read in all particles in the simulation and bin them into 85 logarithmically spaced spherical bins between $0.01R_{\rm200m}$ and $5R_{\rm200m}$.
The dark matter, gas or total mass density profiles are computed by summing the mass of the dark matter particles, gas cells and stellar particles within each radial bin and dividing by the bin volume.

We also compute a galaxy number density profile.
Galaxies in this work are as defined in Section \ref{sec:methods_samples}: subhaloes with $M>10^9\rm{M}_{\odot}$ and a non-zero stellar mass.
There are fewer galaxies in each halo than particles and cells, so we instead use 41 linearly spaced bins between 0 and $5R_{\rm200m}$.
This ensures that there are an adequate number of galaxies in the inner bins while maintaining sufficient resolution in the outer regions to identify the splashback feature.
The density profile is then the number of galaxies in each bin divided by the bin's volume.

Individual haloes are not themselves spherically symmetric objects and the averaged density profiles can be noisy, especially for low-mass haloes.
Therefore, we stack sets of haloes with similar masses or accretion rates.
When grouping by mass, we separate the haloes into four logarithmically spaced mass bins, with $\log_{10}\left(M_{\rm200m}/{\rm M}_{\odot}\right)$ between $13-13.5, 13.5-14, 14-14.5,$ or $14.5-15$, respectively.
For accretion rate, we separate haloes into eight bins with $\Gamma$ between $0-0.5, 0.5-1, 1-1.5, 1.5-2, 2-2.5, 2.5-3, 3-4,$ or $4-5$, respectively.
The number of haloes in each mass or accretion rate bin for TNG300-1 and TNG300-1-DM is shown for a few representative redshifts in Table \ref{table:sample_size}.
The number of haloes in each mass or accretion rate bin for the lower resolution runs is shown in Table \ref{table:res_sample_size} in Appendix \ref{apx:convergence}.
Radial bins for each halo are normalised by $R_{\rm200m}$.
We take the median average density value in each radial bin to yield a stacked density profile as a function of $R_{\rm200m}$.

To identify the splashback feature, we fit our density profiles with the analytic function proposed in \citet{Diemer2014}:
\begin{align}
\label{eq:density}
    \indent\rho(r) &= \rho_{\rm{inner}} \times f_{\rm{trans}} + \rho_{\rm{outer}} \\
    \indent\rho_{\rm{inner}} &= \rho_{\rm{Einasto}} = \rho_{\rm{s}} \exp{\left( -\frac{2}{\alpha} \left[\left(\frac{r}{r_{\rm{s}}}\right)^{\alpha} - 1 \right] \right)}  \nonumber \\
    \indent f_{\rm{trans}} &=  \left[ 1 + \left(\frac{r}{r_{\rm{t}}}\right)^\beta \right]^{-\frac{\gamma}{\beta}} \nonumber \\
    \indent\rho_{\rm{outer}} &= \rho_{\rm{m}} \left[b_e \left( \frac{r}{5R_{200\rm{m}}} \right)^{-S_e} +1 \right]\:. \nonumber
\end{align}
The parameters $\rho_s, r_s, r_t, \alpha, \beta, \gamma, b_e,$ and $S_e$ are left free to vary for a given stacked density profile.
This formula combines descriptions for the inner region of a halo $\rho_{\rm inner}$, in this case an Einasto profile, and for the outer region $\rho_{\rm outer}$, where the profile begins to flatten to the mean density of the universe.
The transitional region between the inner and outer regions is described by $f_{\rm trans}$.

This formula was developed for dark matter only simulations, and we have found that it continues to work well for hydrodynamic simulations.
To fit the gas profile, which does not follow an Einasto profile at low radius due to pressure, we adapt Equation 3 from \citet{Vikhlinin2006} to adjust the inner region such that
\begin{equation}
    \rho_{\rm inner,gas}\! =\! \rho_s {\left(\! \frac{r}{r_s} \!\right)^{\! -\!\frac{\alpha}{2}}}\! {\left(\! 1\!+\!\left(\frac{r}{r_s}\right)^{\!2} \!\right)^{\! -\!\frac{3b}{2}+\frac{\alpha}{4}}}\! {\left(\! 1\!+\!\left(\frac{r}{r_x}\right)^{\!3} \!\right)^{\! -\!\frac{\epsilon}{6}}}\!,\!
    \label{eq:gas_density}
\end{equation}
where $\rho_s, r_s, r_x, \alpha, b,$ and $\epsilon$ are left free to vary when fitting the density profile.
Since the gas density consists only of baryons, we also adjust $\rho_{\rm m}$ in the $\rho_{\rm outer}$ component of the density profile to use the Universal baryon fraction: $\frac{\Omega_{\rm b}}{\Omega{\rm m}}\rho_{\rm m}$.

\subsubsection{Fitting methods}
\label{sec:methods_fitting}

\begin{figure}
    \centering
    \includegraphics[width=\linewidth]{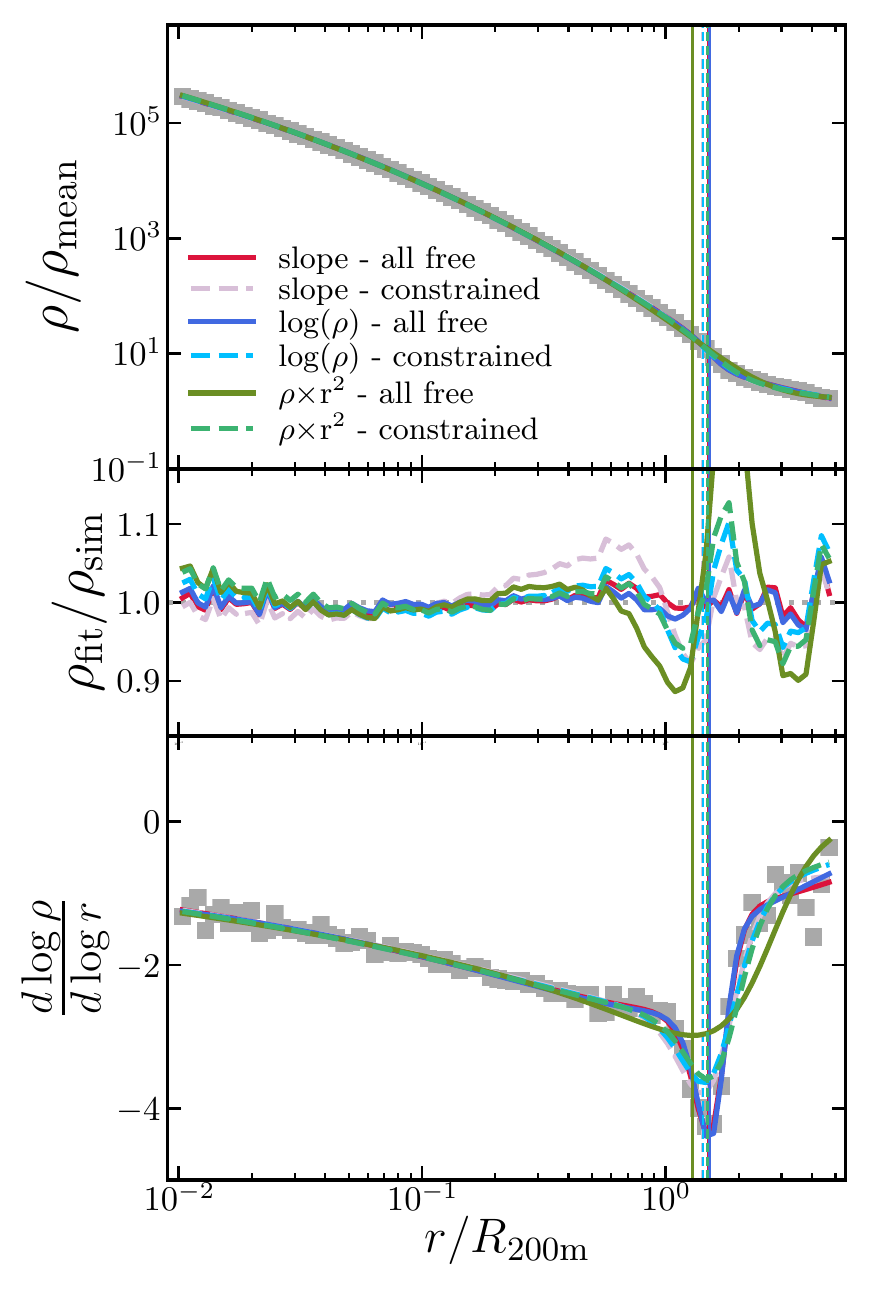}
    \vspace{-.3cm}
    \caption{The stacked density profile from TNG300-1-DM for haloes with $10^{14.5}{\rm M}_{\odot} < M_{\rm200m} < 10^{15}{\rm M}_{\odot}$ at $z=0$.  We test six fitting methods using Equation \ref{eq:density} or \ref{eq:density_derivative}.  The red lines fit the logarithmic slope directly using Equation \ref{eq:density_derivative}, the blue lines fit the profile using the log of Equation \ref{eq:density} and the green lines fit the profile using Equation \ref{eq:density} multiplied by $r^2$.  The darker solid lines leave all eight parameters in the fitting function free, while the lighter dashed lines constrain three parameters according to empirical trends and use five free parameters.  The vertical lines show $R_{\rm sp}$ for each fitting method found by identifying the minimum of the analytic derivative.  Each fit follows the sharp decline in the slope to a varying degree, and there is a small difference in the resulting value for the splashback radius.}
    \label{fig:fit_comp}
\end{figure}

\begin{figure*}
    \centering
    \includegraphics[width=\linewidth]{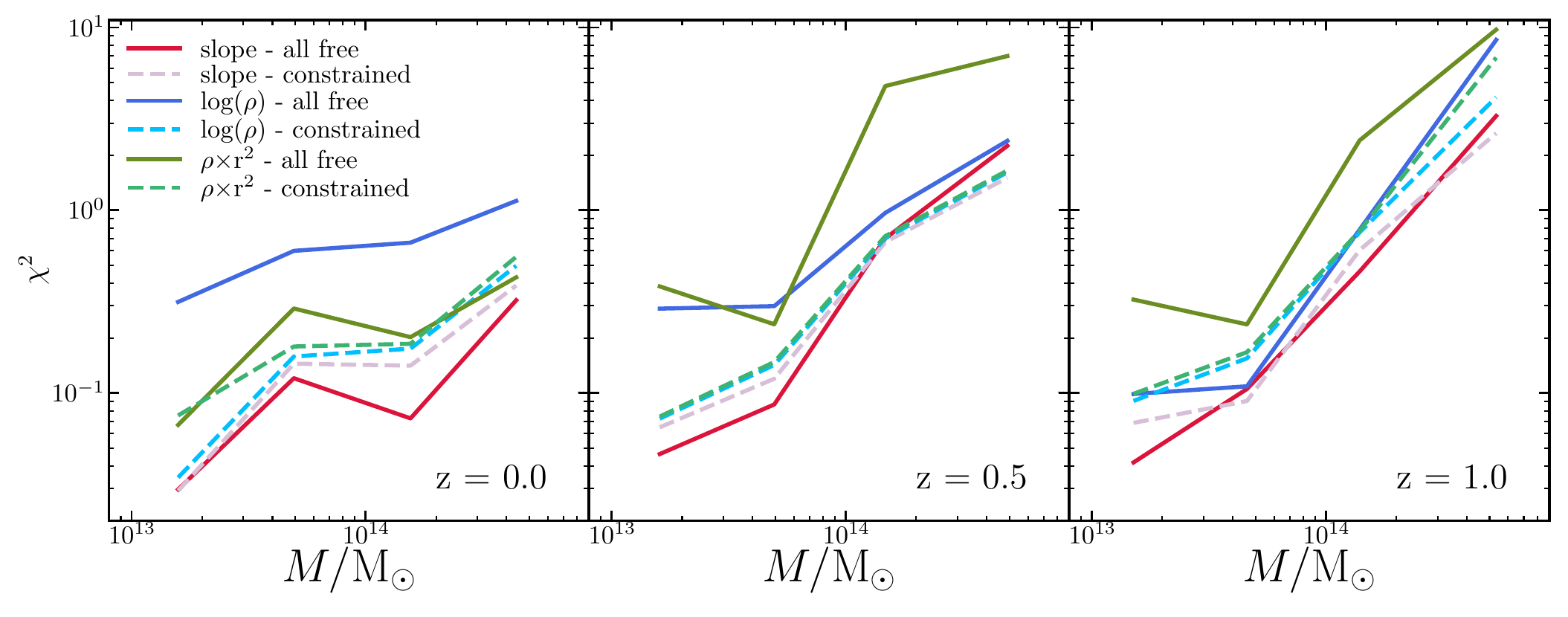}
    \includegraphics[width=\linewidth]{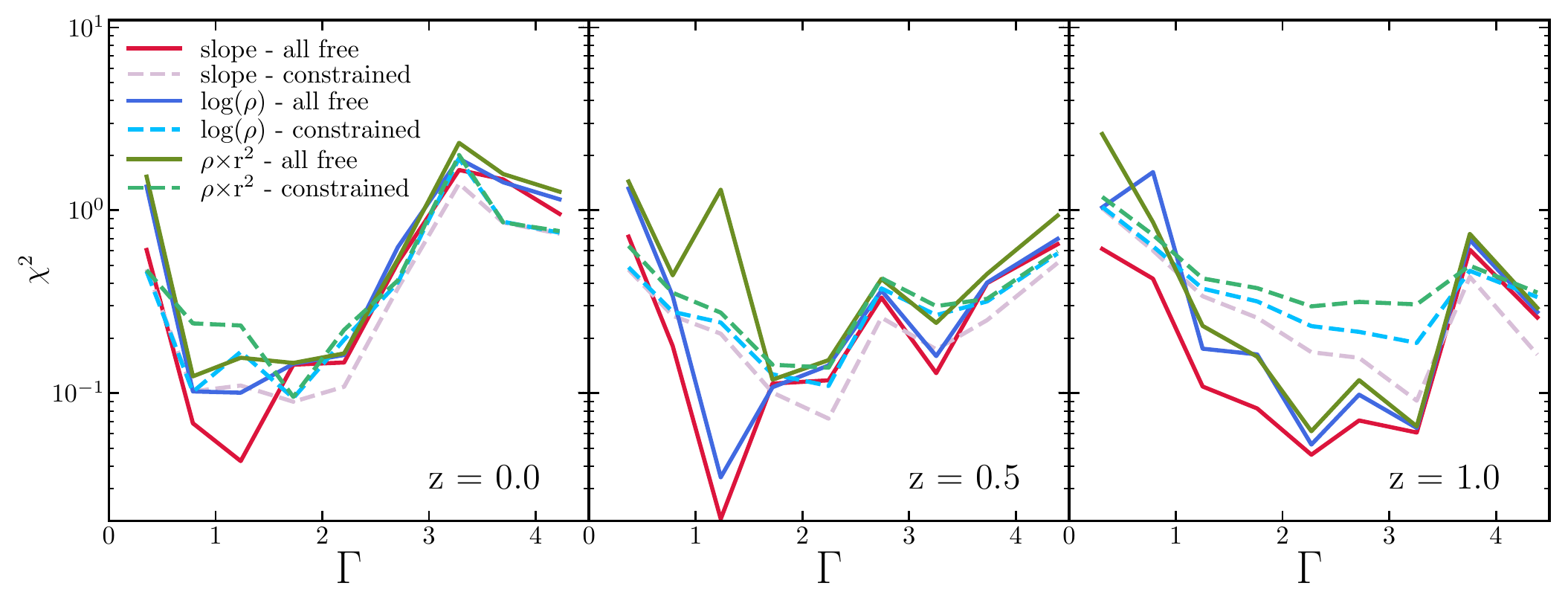}
    \vspace{-.3cm}
    \caption{The $\chi^2$ value defined as in Equation \ref{eq:chi2} as a function of halo mass (top) or accretion rate (bottom) for redshifts 0, 0.5 and 1 (left, middle, and right panels).  The difference between the numerical logarithmic derivative and the analytic logarithmic derivative is computed for points between $0.8R_{\rm200m}$ and $2R_{\rm200m}$, which encloses the splashback region for all our profiles.  The red lines show $\chi^2$ obtained from minimising the fit to the slope, the blue lines are obtained from minimising $\log(\rho)$ and the green lines are obtained from minimising $\rho\times r^2$.  Solid lines leave all parameters from Equation \ref{eq:density} free, and dashed lines fix $\alpha, \beta,$ and $\gamma$.  The solid red line, fitting the slope with all parameters free, produces consistent fits across haloes and redshifts and is the method we use for the remainder of our results.}
    \label{fig:chi2}
\end{figure*}

\begin{figure*}
    \centering
    \includegraphics[width=\linewidth]{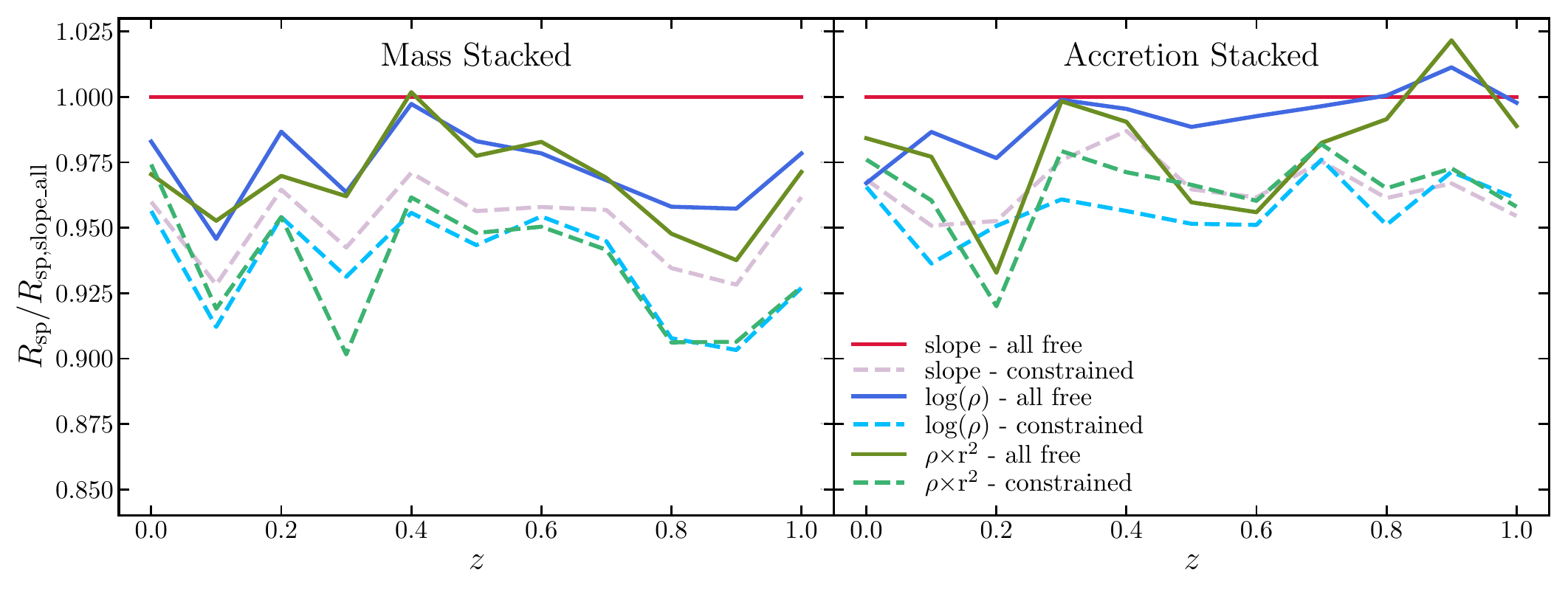}
    \vspace{-.3cm}
    \caption{The fractional difference of $R_{\rm sp}$ for each of the six fiting methods as compared to fitting the slope with all eight free parameters.  We show the fractional difference for density profiles stacked by mass (left) and by accretion rate (right).  We fit the density profile for each redshift and mass or accretion rate range for each method.  We then take the median $R_{\rm sp}$ value for the set of mass or accretion rate ranges for each redshift and fitting method.  The slope fit with all parameters free is generally higher than using other fitting methods on the same density profile by $\sim5\%$.}
    \label{fig:avg_fractional_fit}
\end{figure*}

In addition to exploring the variation of $R_{\rm sp}$ with mass, accretion rate, redshift, and halo components, we also explore the impact of the method used to fit Equation \ref{eq:density} to the stacked profiles.
We compute $R_{\rm sp}$ by identifying the minimum of the logarithmic derivative, so having a robust, accurate fitting method is critical to obtaining reliable results.
Previous work has used a variety of methods to characterise the density profile and find the steepest slope.
\citet{Diemer2014} computed the profile slope using a Savitsky-Golay filter, which provides a piecewise polynomial fit for a moving window of points along the profile, although they did not identify $R_{\rm sp}$ as the minimum of the slope.
Others, e.g. \citet{More2015} and \citet{More2016}, fit the profile using Equation \ref{eq:density} with a reduced number of free parameters and found the steepest slope from the derivatives of their fit.
Observational work, e.g. \citet{Baxter2017} and \citet{Chang2018}, used a Bayesian analysis to fit the parameters to projected density profiles of galaxies.

Several of the fit parameters can be constrained due to known relations or degeneracies, as in e.g. \citet{Diemer2014} and \citet{More2015}.
The Einasto parameter $\alpha$ in Equation \ref{eq:density} can be set according to Equation 5 from \citet{Gao2008} for dark matter haloes:
\begin{equation}
    \label{Gao}
    \indent \alpha = 0.155+0.0095\nu^2\:,
\end{equation}
where $\nu$ is the peak height of the haloes.
The peak height is related to the size of the overdensity from which the halo formed and has a one-to-one correspondence with halo mass and redshift.
To use this in our fits of stacked haloes, we take the median mass of haloes in the sample and convert to peak height for the given redshift.
Two additional parameters, $\beta$ and $\gamma$ may also be fixed.
\citet{Diemer2014} find that $\beta=4$ and $\gamma=8$ provide a good fit if haloes are stacked by mass, while $\beta=6$ and $\gamma=4$ provide a good fit if haloes are stacked by accretion rate.

Previous work used the best-fit parameters of the density profile then differentiated to find the minimum of the slope.
However, as we are interested in the splashback feature that manifests itself as the minimum of the slope, we propose fitting the slope directly by taking the numerical logarithmic derivative of the stacked halo profiles and the logarithmic derivative of Equation \ref{eq:density}.
We compute the numerical derivative of the density profiles using a central finite difference scheme with fourth-order accuracy for the logarithmic profile:
\begin{equation}
    \label{eq:num_deriv}
    \begin{aligned}
    &\frac{d\log\rho}{d\log r}(r_0) = \\ &\frac{\frac{1}{12}\!\log\rho(r_{\!-\!2}) - \frac{2}{3}\!\log\rho(r_{\!-\!1}) + \frac{2}{3}\!\log\rho(r_{\!+\!1}) - \frac{1}{12}\!\log\rho(r_{\!+\!2})}{\log r_{\!+\!2}-\log r_{\!-\!2}}
    \end{aligned}
\end{equation}
where $r_0$ is the point for which the derivative is being calculated and $r_{\pm1}, r_{\pm2}$ are the surrounding points.
The logarithmic derivative of Equation \ref{eq:density} is given by:
\begin{align}
    \label{eq:density_derivative}
    \indent & \frac{d\log{\rho}}{d\log{r}} = \frac{r}{\rho} \frac{d\rho}{d r} \\
    {\rm with}\nonumber\\
    &\frac{d\rho}{d r} = \frac{d\rho_{\rm inner}}{dr} \times f_{\rm trans} + \rho_{\rm inner} \times \frac{df_{\rm trans}}{dr} + \frac{d\rho_{\rm outer}}{dr}\:, \nonumber
\end{align}
where $\rho_{\rm inner}$, $f_{\rm trans}$, and $\rho_{\rm outer}$ are given in Equation \ref{eq:density} with derivatives
\begin{align}
    \label{eq:function_derivatives}
    \indent\frac{d\rho_{\rm inner}}{dr} &=  -\frac{2}{r_s} \left(\frac{r}{r_s}\right)^{\alpha-1} \times \rho_{\rm inner}  \nonumber \\
    \frac{df_{\rm{trans}}}{dr} &=  \left(1 + \left(\frac{r}{r_t}\right)^{\beta}\right)^{-\frac{\gamma}{\beta} - 1} \left(-\frac{\gamma}{r_t}\right) \left(\frac{r}{r_t}\right)^{\beta - 1} \\
    \frac{d\rho_{\rm{outer}}}{dr} &= -\frac{\rho_{\rm m} b_e s_e}{5R_{\rm 200m}} \left(\frac{r}{5R_{\rm 200m}}\right)^{-s_e - 1}\:. \nonumber
\end{align}

When fitting the density profiles of the gas, we adjust the inner portion using Equation \ref{eq:gas_density} with the derivative
\begin{align}
    \label{eq:gas_derivative}
    \frac{d\rho_{\rm{inner,gas}}}{dr} = &-\! \frac{\rho_s}{r_s} \frac{\alpha}{2} \!\times\! \left[ 1\! +\! \left(\!\frac{r}{r_s}\!\right)^{\!2} \right]^{ -\!\frac{3b}{2} + \frac{\alpha}{4} } \!\!\!\!\!\times\! \left[ 1\! +\! \left(\!\frac{r}{r_x}\!\right)^{\!3} \right]^{-\!\frac{\epsilon}{6}} \nonumber\\
    &+ \frac{2r}{r_s^2} \left(-\frac{3b}{2} + \frac{\alpha}{4}\right) \left[ 1 + \left(\frac{r}{r_s}\right)^2 \right]^{-\frac{3b}{2} + \frac{\alpha}{4} - 1} \nonumber\\
    &\indent\times \rho_s  \left(\frac{r}{r_s}\right)^{-\frac{\alpha}{2}} \times \left[ 1 + \left(\frac{r}{r_x}\right)^3 \right]^{-\frac{\epsilon}{6}} \nonumber\\
    &- \frac{\epsilon}{2r_x} \left(\frac{r}{r_x}\right)^2 \left[1 + \left(\frac{r}{r_x}\right)^3 \right]^{-\frac{\epsilon}{6} - 1.0} \nonumber\\
    &\indent\times \rho_s  \left(\frac{r}{r_s}\right)^{-\frac{\alpha}{2}} \times \left[ 1 + \left(\frac{r}{r_s}\right)^2 \right]^{ -\frac{3b}{2} + \frac{\alpha}{4}}\!.
\end{align}

Once the slope is fit, we fit an additional normalisation parameter $N$:
\begin{equation}
    \label{eq:density_norm}
    \indent \rho(r) = N \rho_{\rm fit}\:.
\end{equation}
This adjusts the normalisation of the density profile on the logarithmic plot without changing the slope, and we use it to show the results of our fits in the figures presented in this paper.
This normalisation parameter is not used in any other calculations.

We use a least squares method to fit the analytic function to our density profiles then identify the minimum of $\frac{d\log\rho}{d\log r}$ to calculate the value of $R_{\rm sp}$.
We test the impact of the fitting scheme on $R_{\rm sp}$ by varying the weighting used in the least squares fit.
For the weighting schemes, we minimise the difference in $\rho\times r^2$, $\log\rho$, or $\frac{d\log\rho}{d\log r}$.
When fitting $\rho\times r^2$ or $\log\rho$, we perform the fit, then take the derivative to calculate $R_{\rm sp}$.  When fitting $\frac{d\log \rho(r)}{d\log r}$, we use the method described above.

We also test the impact of fixing parameters or leaving them free to vary.  For each of these weightings, we perform fits with all eight parameters given in Equation \ref{eq:density} free or fixing $\alpha, \beta,$ and $\gamma$ as described above.
This gives us a total of six fitting schemes.
An example of each of these fits on the same density profile is shown in Figure \ref{fig:fit_comp}.  Previous work, e.g. \citet{Diemer2014}, \citet{Diemer2020a} and
\citet{More2015}, used the constrained $\rho r^2$ weighting to fit their profiles.

Our method for finding $R_{\rm sp}$ for a density profile can be summarised as follows:
\begin{itemize}
    \item Calculate the spherically averaged density profiles of individual haloes out to $5R_{\rm200m}$.
    \item Select a set of haloes within a given mass or accretion rate range and take the median density values.
    \item Use a least squares method to find the best-fit parameters.
    \item If constraining some parameters to leave five free, fix $\alpha$ according to Equation \ref{Gao}.  Set $\beta$ and $\gamma$ to 4 and 8 if haloes are stacked by mass or to 6 and 4 if haloes are stacked by accretion rate.
    \item Use the best-fit parameters to calculate the minimum of the analytic derivative of the density profile.
\end{itemize}

\subsubsection{Error estimation}
\label{sec:methods_error}
We use the bootstrapping method to estimate the uncertainty in our calculation of $R_{\rm sp}$.
We randomly sample our halo profiles with replacement 10,000 times.
This sampling occurs before separation by mass or accretion rate, so we create a sample of all haloes and then separate them into sets of stacked haloes.
If a sample results in zero haloes in a given range, e.g. no high mass haloes, we discard that sample and create another.
We compute independent samples for each redshift, resolution and halo component.
We calculate the splashback radius of the stacked profiles for haloes in each sample and show the 16th and 84th percentiles, approximating a standard deviation of a Gaussian distribution.

\begin{figure*}
    \centering
    \includegraphics[width=\linewidth]{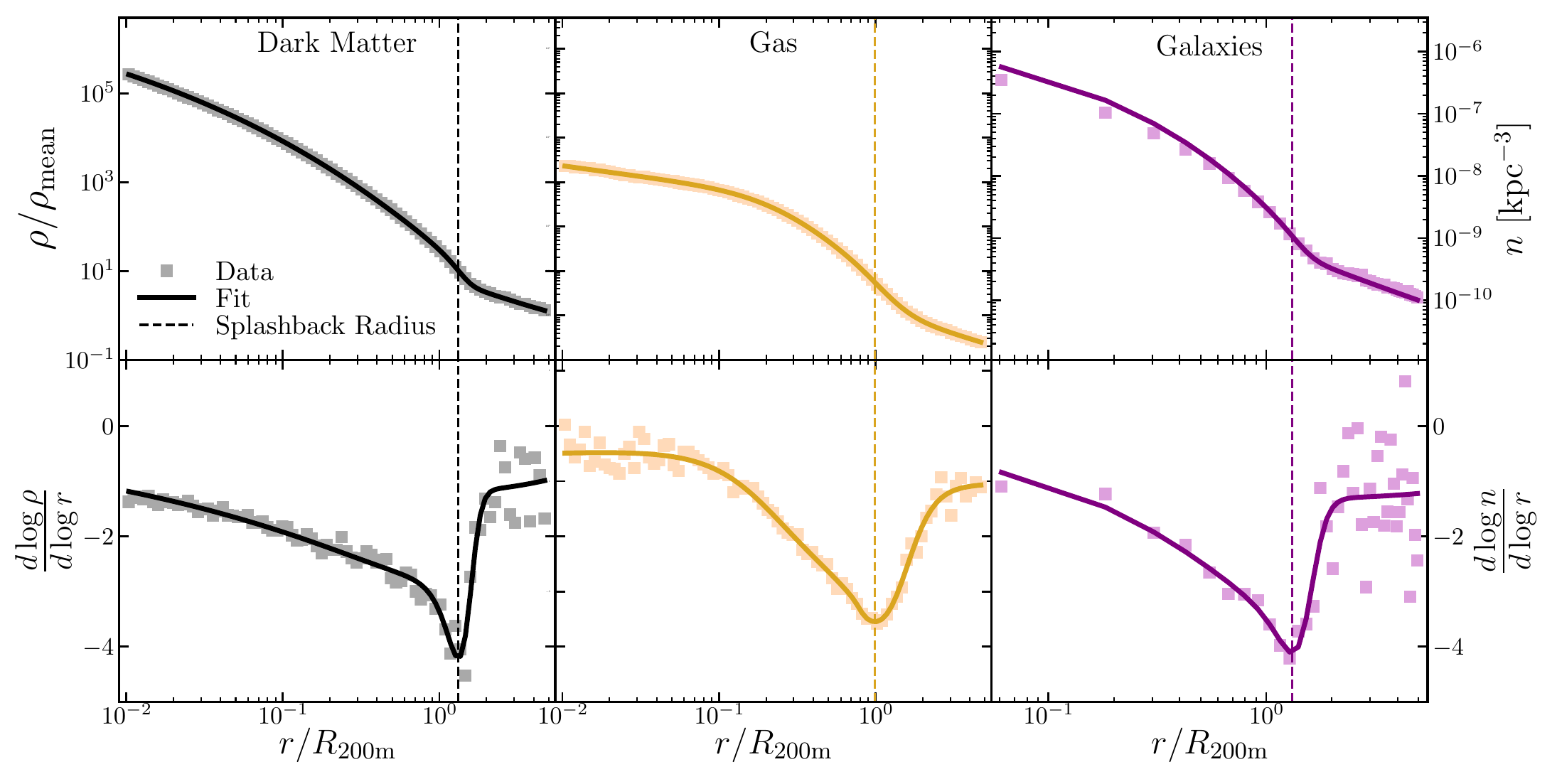}
    \vspace{-.3cm}
    \caption{Examples of stacked density profiles for haloes between $10^{14}\rm{M}_{\odot}$ and $10^{14.5}\rm{M}_{\odot}$ at $z=0$ in TNG300-1.  The calculated densities and fits for dark matter are on the left (black), gas in the middle (orange) and galaxy number density on the right (purple).  The lighter squares are the densities and slopes calculated from the simulation, the solid lines are the fit and dashed vertical lines show the location of the splashback radius for each density profile.  The fits use Equation \ref{eq:density}, adjusted for the gas profile using Equation \ref{eq:gas_density}.  These equations provide a good description for the density profile for each component.}
    \label{fig:stacked_density_examples}
\end{figure*}

\section{Results}
\label{sec:results}
In this section, we explore the splashback feature in a sample of haloes selected from the TNG300 volume.
Appendix \ref{apx:convergence} utilises the lower resolution simulations to highlight the numerical convergence of our results.
Throughout this section, we compare to the results in \citet{More2015} and \citet{Diemer2020a}.
They use \textit{N}-body dark matter only simulations to build a model that predicts the location of $R_{\rm sp}$ for a given halo.
These models are implemented in the \textsc{Python} toolkit \textsc{Colossus} \citep{Diemer2018}, yielding predicted $R_{\rm sp}\,/\,R_{\rm200m}$ values.
For \citet{More2015}, the prediction is given as a function of accretion rate and redshift.
The fitting function for $R_{\rm sp}$ from \citet{Diemer2020a} is implemented as a function of redshift, accretion rate and peak height.
These models can also be used as functions of mass by assuming relations between accretion rate and mass.
Additionally, \textsc{Colossus} provides a convenient function to convert between peak height and halo mass at a given redshift for a given cosmology.

\citet{More2015} explored the splashback radius and enclosed mass as a function of halo mass, accretion rate, and peak height for dark matter haloes.
They defined the splashback radius as the steepest slope of the spherically-averaged stacked median density profile, computed by fitting the density profile.
\citet{Diemer2020a} studied the splashback radius using the orbital dynamics of dark matter particles.
They compute the apocentre of first orbit by tracing particle trajectories as they fall into the halo.
As each particle reaches a slightly different apocentre, the splashback radius is approximated as the distance enclosing a percentile of the infalling particles.
Although there is no exact match between any given percentile and the point of steepest slope found in \citet{More2015}, we use the 75th percentile of the \citet{Diemer2020a} as a basis of comparison.

\subsection{Impact of the fitting method}
\label{sec:results_fitting}
We begin by examining the impact of the fitting method on the recovered $R_{\rm sp}$.
Using the six weightings described in Section \ref{sec:methods_fitting}, we fit stacked profiles extracted from the highest resolution dark matter only simulation.
Additionally, we examine the impact of leaving all eight parameters in Equation \ref{eq:density} free, ``all free'', or constraining the $\alpha, \beta,$ and $\gamma$ parameters, ``constrained''.

The aim of fitting the profile is to accurately identify the sharp decrease in the slope of the profile.
As a measure of how well fit the profile is, we calculate a $\chi^2$ value defined as follows:
\begin{equation}
    \indent\chi^2 = \sum\frac{\left(\frac{d\log\rho_{\rm fit}}{d\log r} - \frac{d\log\rho_{\rm sim}}{d\log r}\right)^2}{N-m}\:,
    \label{eq:chi2}
\end{equation}
where $\rho_{\rm fit}$ is the analytic profile fit and $\rho_{\rm sim}$ is the density profile obtained via the method outlined in Section \ref{sec:methods_density}. 
$N$ is the number of data points we use to calculate $\chi^2$ and $m$ is the number of free parameters in the fit.
To focus on $R_{\mathrm{sp}}$, we only include points in the range $0.8\leq R/R_{\rm200m}\leq2$, although the fit is performed using all points in the density profile.
This also increases the significance of $m$ compared to $N$ in the denominator of Equation \ref{eq:chi2} so we can better compare fits with different numbers of free parameters. However, calculating $\chi^2$ using all points in the fit does not significantly change our results since the difference between the fit and numerical results is dominated by the splashback region.

This $\chi^2$ value is shown for each fit as a function of halo mass and accretion rate at $z=0$, $0.5$ and $1$ in Figure \ref{fig:chi2}.
Fitting to the density derivative with all eight parameters free, shown by the solid red line, consistently yields a lower $\chi^2$ across our mass, accretion rate and redshift ranges.
The increase of $\chi^2$ with mass is due to having fewer haloes with larger mass, which results in more scatter in the median profile and therefore a larger $\chi^2$.

In addition to having a comparable $\chi^2$ value as previously used methods, we found that fitting the slope worked more robustly on larger variety of profiles.  This resulted in less tuning of initial guesses in the fitting routines and fewer failed fits.  This was especially important for the gas profiles, which have more shape variation than the dark matter profiles, and galaxy profiles, which are noisier than the dark matter profiles.  Qualitatively, fitting the slope with all parameters free consistently fit the sharp decrease in all profiles we fit while the other methods sometimes did not sufficiently decrease, as in e.g. the $\rho\times r^2$ method in Figure \ref{fig:fit_comp}.

How well a method fits the sharp decrease in the derivative of the profile systematically influences the $R_{\rm sp}$ recovered.
Figure \ref{fig:avg_fractional_fit} shows the fractional difference between $R_{\rm sp}$ obtained for all fitting methods relative to the eight free parameter density derivative fit.
At each redshift, we calculate the median fractional difference of fits for each mass (left) or accretion rate (right) range and plot the result.
Fitting the density derivative with eight free parameters yields $R_{\mathrm{\rm sp}}$ values that are $\sim5\%$ larger than the other methods.
Those fitting methods that fail to capture the sharp drop in density around the splashback feature are generally biased low relative to the minimum of the derivative, as shown in Figure \ref{fig:fit_comp}.

More accurately capturing the derivative of the density profile about the point of maximal change appears to result in the recovery a systematically larger splashback radius.
For the remainder of the paper, all results are derived from fitting the derivative of the density profile with all parameters left free to vary, as it more accurately captures the derivative of the density profile.

\subsection{The splashback radius of halo components}
\label{sec:results_components}
We now explore three components of the haloes in the TNG300 simulations: dark matter density, gas density and galaxy number density.
The galaxy profiles are the number densities of all galaxies around a halo, where a galaxy is defined in Section \ref{sec:methods_samples}.

Figure \ref{fig:stacked_density_examples} shows example stacked density profiles and best-fit models for the dark matter, gas and galaxy profiles of haloes in TNG300-1 at $z=0$ for haloes in the range $10^{14}{\rm M}_{\odot}\leq M<10^{14.5}{\rm M}_{\odot}$.
The top panel shows the median stacked density profiles, while the bottom panel shows the corresponding logarithmic slope.
The splashback radius is shown by the vertical line, which is defined as the minimum of the slope.

Both galaxies and dark matter are primarily influenced by orbital dynamics, resulting in splashback, but the gas is highly collisional and therefore governed by shocks.  While both splashback and shocks lead to a drop in density, this also results in the wider, shallower steepening in the gas profiles compared to the dark matter and galaxy profiles.

In idealised spherical collapse models, e.g. \citet{Shi2016b}, the accretion shock position is expected to be near the splashback radius.
It is also expected to behave similarly to the splashback radius and decreases with accretion rate, although this similarity is coincidental for gas with an adiabatic constant $\gamma\approx5/3$ (also see \citet{Bertschinger1985}).
However, the presence of mergers in the accretion history of halos can significantly change the behaviour of shocks.

In more realistic simulations, e.g. \citet{Lau2015} and \citet{Aung2020}, the accretion shock occurs at $\sim1.5-2$ times the splashback radius.
Recent analyses have shown that a number of physical effects can cause the decreases in the gas profiles, including accretion shocks, runaway merger shocks, and contact discontinuities \citep{Aung2020, Zhang2021}.
Although \citet{Aung2020} find that the accretion shock occurs at larger radius than the splashback radius and the dip in the gas density profile, the accretion shock radius also decreases with halo accretion rate similarly to the splashback radius.
Contact discontinuities where the runaway shock overcomes the accretion shock could contribute to a drop in gas density near the virial radius that is shallower compared to the splashback feature \citep[e.g.][]{Zhang2021}.
These contact discontinuities can easily be formed during mergers due to stripping from the subcluster \citep[see][for review]{Markevitch2007} and can persist for $3-5$ Gyr \citep{Zhang2021}.

Therefore, we expect to find the steepening in the gas density profile in a similar place as in the dark matter.
However, some difference in its location and the behaviour of the density profile is to be expected.
Understanding these similarities and differences could provide an essential basis for the detection of the splashback feature in observational data.
Exact observational predictions, however, are beyond the scope of this paper and we leave it for future work.

\subsubsection{Dependence on halo mass}
\label{sec:results_mass}
We examine the position of $R_{\rm sp}$ as a function of halo mass for each halo component by splitting the haloes into four mass bins between $10^{13}\rm{M}_{\odot}$ and $10^{15}\rm{M}_{\odot}$.
This has been explored in e.g. \citet{Diemer2017b}, although they focused primarily on the correlation between the splashback radius and halo accretion rate.  Halo mass, however, is more easily inferred from observations than accretion rate.

\begin{figure}
    \centering
    \includegraphics[width=\linewidth]{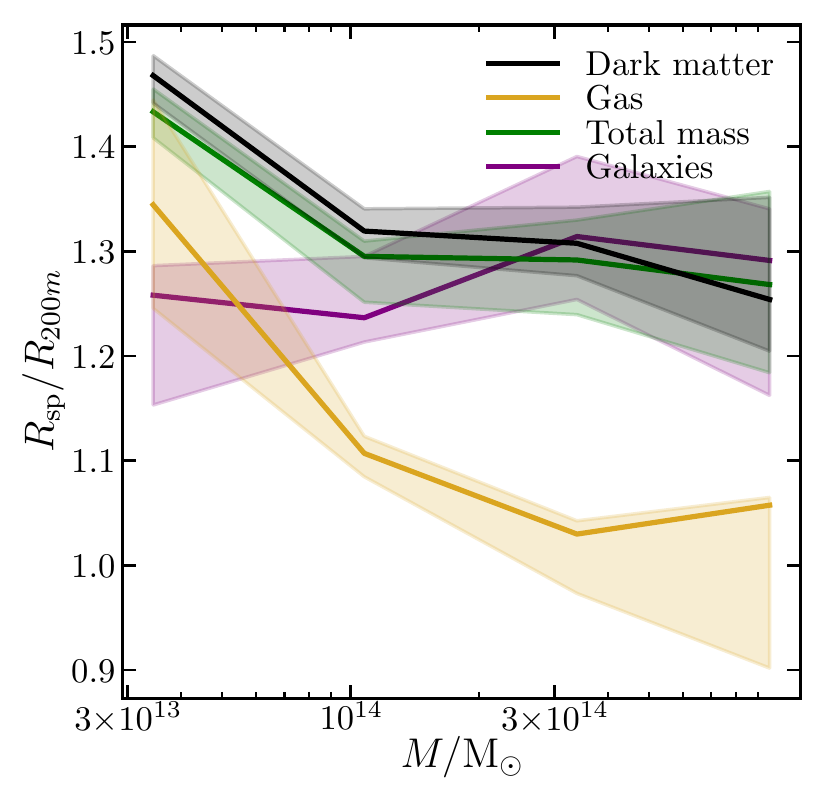}
    \vspace{-.3cm}
    \caption{Location of the splashback radius as a function of halo mass for the dark matter mass, gas mass, total mass and galaxy number density at $z=0$.  We stack the density profiles for haloes with $\log_{10}\left(M_{200\rm{m}}/{\rm M}_{\odot}\right)$ in ${13}-{13.5}$, ${13.5}-{14}$, ${14}-{14.5}$ and ${14.5}-{15}$ and compute the splashback radius of the median profile, shown by the solid lines.  To estimate error, we use a bootstrap method and show the 16th and 84th percentiles as the shaded band around each line.  $R_{\rm sp}$ occurs in a similar location in the total mass profiles as in the dark matter profile.  It also occurs in a similar location for the galaxy profile except for low mass haloes, where dynamical friction is more important.  $R_{\rm sp}$ in the gas profiles tends to occur at a lower radius than in the other profiles.}
    \label{fig:mass_evolution}
\end{figure}

Figure \ref{fig:mass_evolution} shows $R_{\rm sp}/R_{\rm200m}$ calculated as the point of steepest slope as a function of halo mass for the dark matter, gas, total (dark matter, stars and gas) mass and galaxy profiles in the TNG300-1 simulation at $z=0$.
In general, this quantity decreases with increasing halo mass.
The total mass profiles yield results similar to the dark matter profiles.
This is expected because dark matter accounts for most of the mass in haloes.

\begin{figure*}
    \centering
    \includegraphics[width=\linewidth]{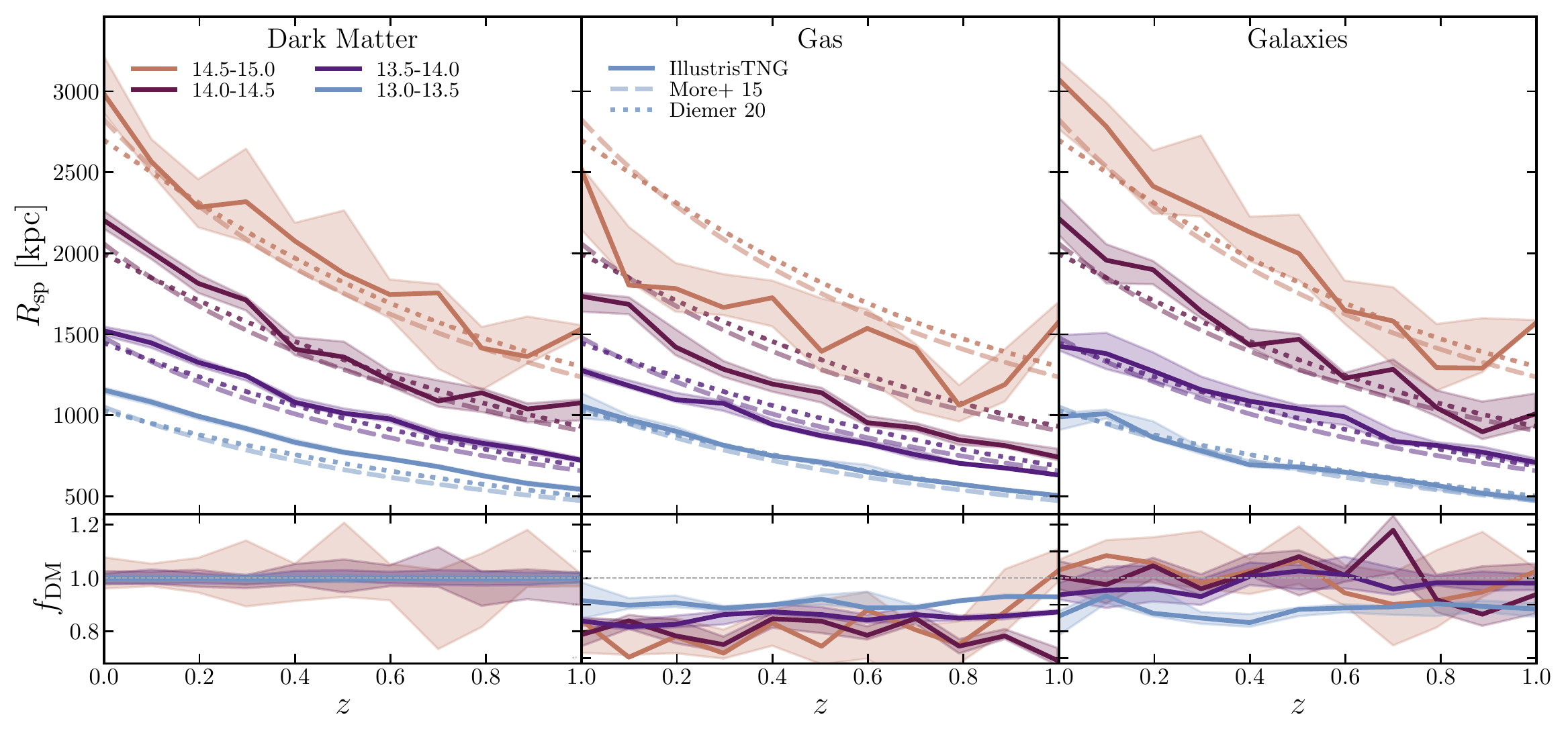}
    \vspace{-.3cm}
    \caption{The top panels show the splashback radius as a function of redshift for the four halo mass bins.  Here we compare results from the dark matter mass, gas mass and galaxy number density profiles.  We stack the density profiles for haloes in four mass bins between $10^{13}\rm{M}_{\odot}$ and $10^{15}\rm{M}_{\odot}$ and compute the splashback radius of the median profile, shown by the solid lines.  Each colour corresponds to a mass bin labeled with the $\log_{10}$ of its bounds.  To estimate error, we use a bootstrap method and show the 16th and 84th percentiles as the shaded band around each line.  We compare to two analytic models, found in \citet{More2015} and \citet{Diemer2020a}, using the middle of each mass range ($\log(M/{\rm M}_{\odot}) = 13.25, 13.75, 14.25$ and $14.75$) at each redshift.  These models predict $R_{\rm sp}$ for dark matter haloes, but we show them in each panel for reference.  Although our normalisation differs, we find a similar evolution of $R_{\rm sp}$ with redshift.  The bottom panels show the fractional difference between each component and the dark matter.  The gas profiles have $R_{\rm sp}$ that is lower than the dark matter $R_{\rm sp}$ across all redshifts by $\sim10-20\%$.  The galaxy profiles have similar $R_{\rm sp}$ as the dark matter and the smallest mass bin consistently produces $R_{\rm sp}$ lower for the galaxies by $\sim10\%$.}
    \label{fig:redshift_evolution_components}
\end{figure*}

\begin{figure*}
    \centering
    \includegraphics[width=\linewidth]{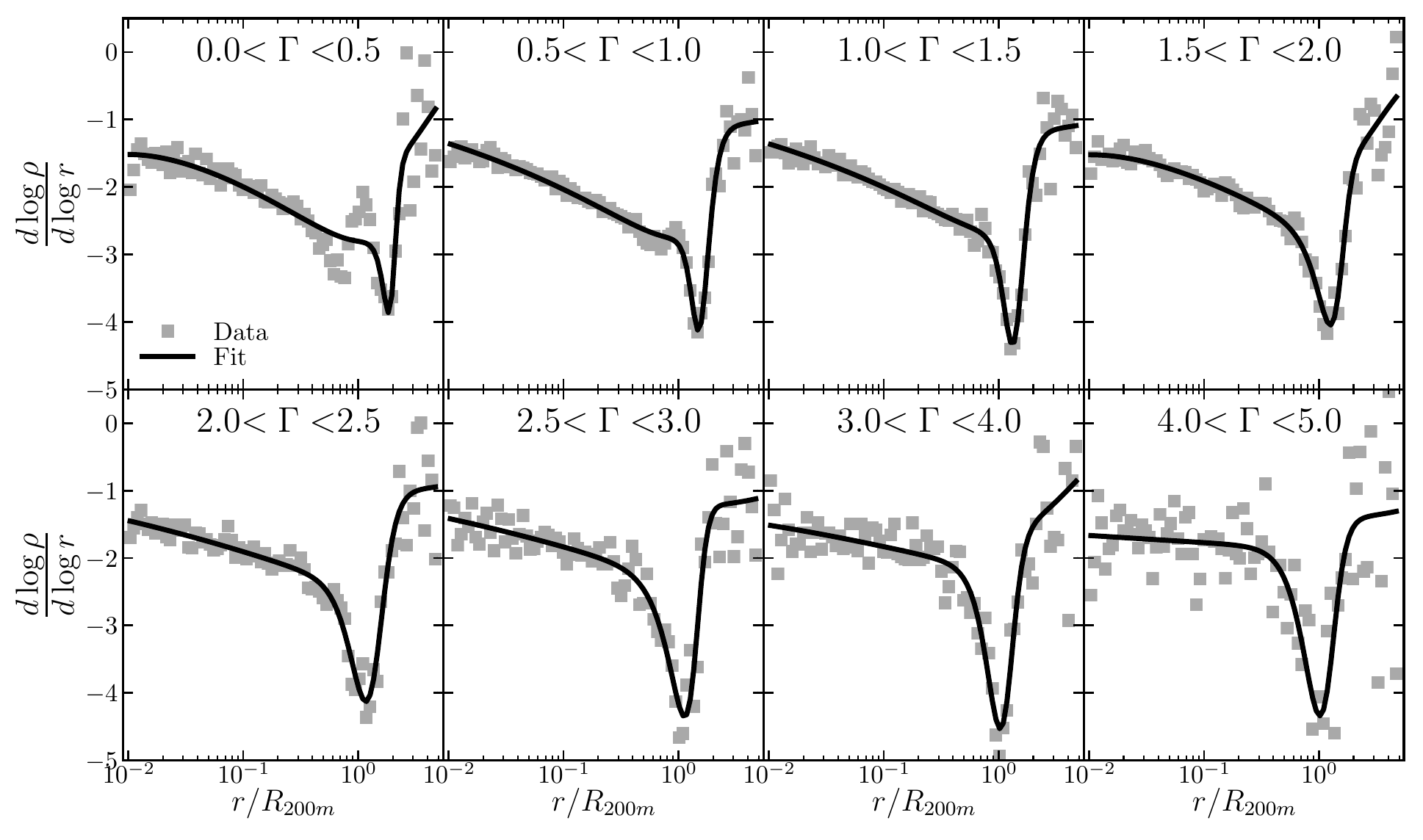}
    \vspace{-.3cm}
    \caption{The logarithmic derivative of the dark matter density profile stacked in each accretion rate range.  The grey points are computed numerically from the simulation, and the solid line is the analytic derivative using the fit described in Section \ref{sec:methods_fitting}.  Lower accretion rates show a second local minimum in the logarithmic derivative, which disappears for $\Gamma>1$.  The fit consistently identifies the minimum due to splashback at slightly larger radius.}
    \label{fig:caustic}
\end{figure*}

The deviation between the steepest slope in the gas and dark matter profiles increases with halo mass.
$R_{\rm sp}/R_{\rm200m}$ is $\sim20\%$ lower for the gas relative to the dark matter for the largest two mass bins, $10^{14-14.5}\rm{M}_{\odot}$ and $10^{14.5-15}\rm{M}_{\odot}$.
In the next largest mass bin, $10^{13.5-14}\rm{M}_{\odot}$, we find that $R_{\rm sp}/R_{\rm200m}$ is $\sim15\%$ lower in the gas than in the dark matter.
Finally, $R_{\rm sp}/R_{\rm200m}$ is $\sim10\%$ lower in the gas relative to the dark matter for the smallest mass bin, $10^{13-13.5}\rm{M}_{\odot}$.

The difference between $R_{\rm sp}$ in the galaxy and dark matter profiles is smaller than between gas and dark matter, but it is more significant for lower mass haloes.
For haloes with masses between $10^{13.5}\rm{M}_{\odot}$ and $10^{14}\rm{M}_{\odot}$, the ratio of $R_{\rm sp}/R_{\rm200m}$ in the galaxy and dark matter profiles is slightly lower than for the largest two mass bins, but all three are in agreement with one.
For haloes with $M_{\rm200m}$ between $10^{13}\rm{M}_{\odot}$ and $10^{13.5}\rm{M}_{\odot}$, $R_{\rm sp}/R_{\rm200m}$ is $\sim12\%$ lower for galaxies than for dark matter.

The tendency for $R_{\rm sp}$ to be lower in the galaxy profile than in the dark matter profile agrees with previous work.
\citet{Deason2020a} found that subhaloes gave a significantly smaller $R_{\rm sp}$ than the dark matter in Local Group simulations.
Notably, the most significant deviation between $R_{\rm sp}$ for galaxies and dark matter occurs for our smallest mass bin, the mass closest to that of the Local Group.
This suggests that there is an environmental effect and that it should not necessarily be assumed that the galaxy number density matches the dark matter mass density.
\citet{Xhakaj2020}, however, found that subhalo profiles yield smaller $R_{\rm sp}$ values even for haloes with $M\sim10^{14}\rm{M}_{\odot}$ but with larger subhaloes of masses $\sim10^{12}\rm{M}_{\odot}$.
We also fit our profiles using different methods, which can produce different values for $R_{\rm sp}$.

\citet{Adhikari2016} noted that subhaloes are subject to dynamical friction as they fall into a halo while dark matter particles are not.
To understand the impact this has on the orbits, they modified the subhalo equation of motion to account for dynamical friction to
\begin{equation}
    \indent\frac{dv_r}{dt} = -\frac{GM(r)}{r^2}-\eta\frac{G^2m_{\rm sub}\rho(r)}{|v_r|^3}v_rf(v_r/\sigma)\:,
    \label{eq:fric_eos}
\end{equation}
where $v_r$ is the radial velocity of the subhalo, $m_{\rm sub}$ is its mass and $M(r)$ is the halo mass enclosed within the orbital radius $r$ with local density $\rho(r)$.
$f(v_r/\sigma)$ is the phase space factor, typically taken to be Maxwellian so $f(x)=\mathrm{erf}(x)-\frac{2}{\sqrt{\pi}}xe^{-x^2/\sigma^2}$.
$\eta$ is a constant that depends on the distribution function of particles and the subhalo's internal structure, which \citet{Adhikari2016} finds to be $\sim1-2$.

To find how the effect of dynamical friction might vary with the mass of the halo, we can assume the velocity of a subhalo is altered over one orbital period $T=2\pi r/v$ by
\begin{equation}
    \indent \Delta v = -\eta\frac{G^2 m_{\rm sub} \rho(r)}{v^2} f T\:.
    \label{eq:delta_v_1}
\end{equation}
For orbital motion $v^2 = \frac{GM(r)}{r}$, we get
\begin{align}
    \indent & 2v\Delta v = -\frac{GM(r)}{r^2}\Delta r\nonumber\\
    \Rightarrow & \Delta v = -\frac{1}{2}v\left(\frac{\Delta r}{r}\right)\:.
    \label{eq:delta_v_2}
\end{align}
Combining Equations \ref{eq:delta_v_1} and \ref{eq:delta_v_2} gives
\begin{equation}
    \indent\frac{\Delta r}{r} = \eta f \frac{4\pi m_{\rm sub}\rho(r)r^3}{M(r)^2}\:.
    \label{eq:delta_r}
\end{equation}

For our density profiles, the density near $R_{\rm200m}$ does not depend strongly on mass and is $\sim20\rho_{\rm m}$.
For $M(r)\sim r^3$, we get a change in radius dependent on halo mass
\begin{equation}
    \indent\frac{\Delta r}{r} \propto \frac{m_{\rm sub}}{M(r)}\:.
    \label{eq:delta_r_prop}
\end{equation}
This implies that dynamical friction has a larger fractional effect in smaller haloes and larger subhaloes as in \citet{Adhikari2016}.
As halo mass increases, this effect becomes negligible and we see the location of $R_{\rm sp}$ more closely agrees with the dark matter results.

Additionally, a subhalo's mass is stripped as it falls into a halo.
The more mass a subhalo loses, the more bound its orbit will become and the position of its first apocentre will decrease.
This further adds to the disparity between the galaxy and dark matter splashback feature.
However, the formation of a galaxy within a subhalo steepens the potential and makes it less susceptible to stripping.
Dark matter only simulations may not fully capture this effect, so it is unclear how well our galaxy number density profiles should agree with dark matter only subhaloes.

\subsubsection{Redshift evolution}
\label{sec:results_redshift}
We now explore the evolution of the splashback feature with redshift.
The redshift evolution of $R_{\rm sp}$ is primarily due to a dependence on $\Omega_{\rm m}$ \citep{Diemer2014,Adhikari2014, More2015, Mansfield2017, Diemer2017b}.
This is easily converted into a redshift dependence assuming the Flat $\Lambda$CDM cosmology of IllustrisTNG.
$R_{\rm sp}$ as a function of redshift for each halo mass bin is shown in Figure \ref{fig:redshift_evolution_components}.
$R_{\rm sp}$ is computed independently for each mass bin and redshift, so haloes at each redshift are treated as independent.
We find that haloes will have a larger $R_{\rm sp}$ than haloes at a higher redshift with the same $M_{\rm200m}$.
At earlier times, in a more dense environment, matter will accrete more quickly onto a halo.
This will increase the mass, and therefore the gravitational potential, of the halo during the first orbit of infalling material, which will decrease the radial orbit of the material \citep{Adhikari2014}.
The gas and galaxy profiles follow a similar evolution with redshift, but they differ from the dark matter profiles in their normalisation and mass dependence as discussed in the previous section.
The trends in mass dependence persist across redshifts for both the gas and galaxy profiles.

\citet{Diemer2017b} explicitly tested the dependence of $R_{\rm sp}/R_{\rm200m}$ on $\Omega_{\rm m}$.
For constant $\Gamma$, $R_{\rm sp}/R_{\rm200m}$ increased, converging for $z\geq2$ where $\Omega_{\rm m}\sim1$.
Similarly, when the cosmology of the simulation is altered, $R_{\rm sp}/R_{\rm200m}$ increases for larger $\Omega_{\rm m}$.
When $\Omega_{\rm m}$ is fixed at 1, $R_{\rm sp}/R_{\rm200m}$ does not evolve with redshift.
The power-spectrum slope also has some impact on this dependence, but this has not been explored.

\begin{figure*}
    \centering
    \includegraphics[width=\linewidth]{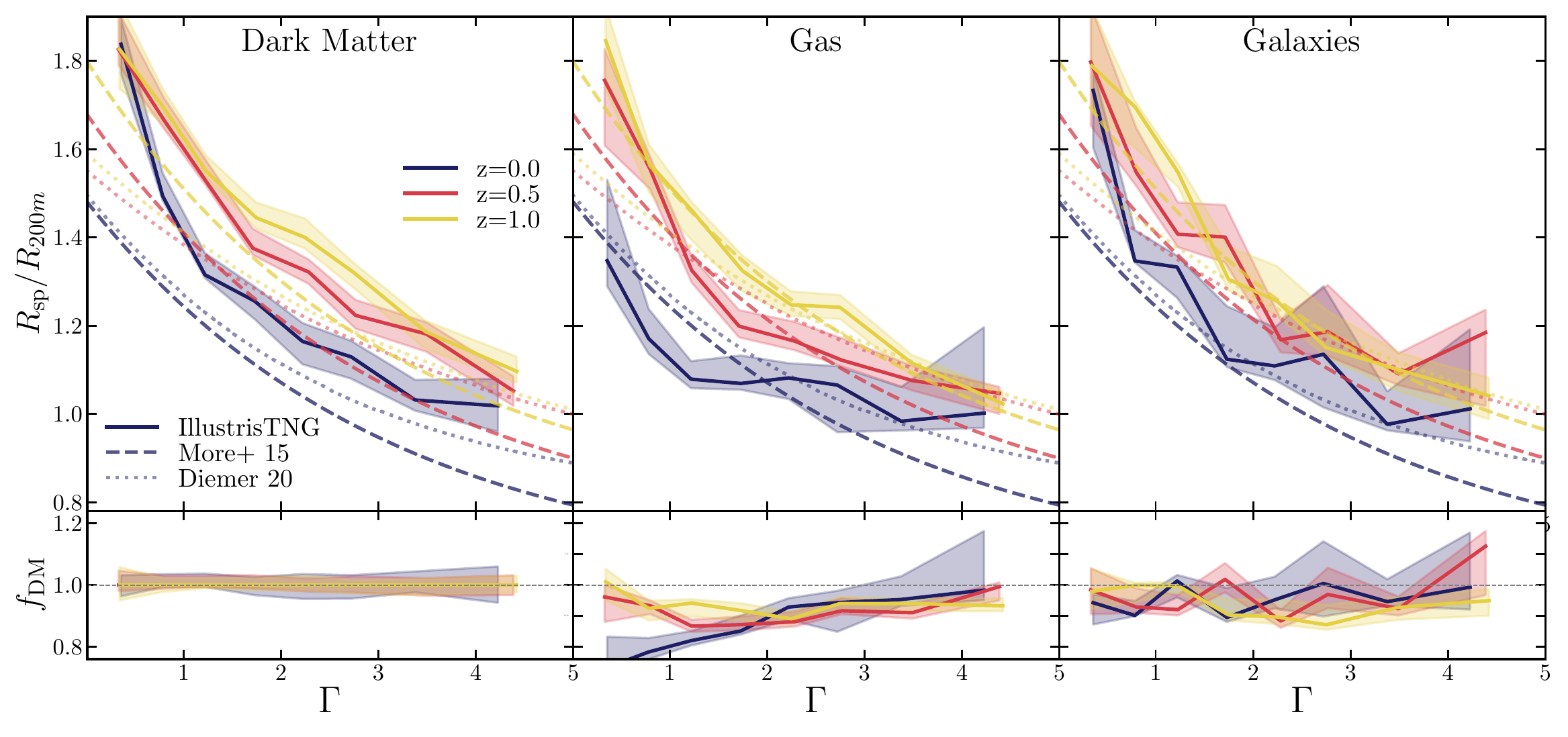}
    \vspace{-.3cm}
    \caption{The top panels show the location of the splashback radius as a function of accretion rate $\Gamma$ for the dark matter mass, gas mass and galaxy number density profiles.  The colours show several redshifts, and we compare to the models described in \citet{More2015} and \citet{Diemer2020a}.  The location of the splashback radius decreases with accretion rate and increases with redshift.  The haloes are stacked based on accretion rates in eight bins between $0$ and $5$.  We compute the splashback radius of the median stacked profile for each accretion range.  To compare to the \citet{More2015} and \citet{Diemer2020a} models, we calculate a value for $R_{\rm sp}/R_{\rm 200m}$ based on the median $R_{\rm200m}$ value for each accretion rate range.  These models predict $R_{\rm sp}$ for dark matter haloes, but we show them in each panel for reference.  The bottom panels show the fractional difference between each component and the dark matter.  The gas profiles result in $R_{\rm sp}/R_{\rm200m}$ lower than the dark matter profiles, especially at lower $\Gamma$, while the galaxy profiles do not deviate significantly from the dark matter profiles.}
    \label{fig:gamma_components}
\end{figure*}

\subsubsection{Dependence on halo accretion rate}
\label{sec:results_accretion}

\begin{figure}
    \centering
    \includegraphics[width=\linewidth]{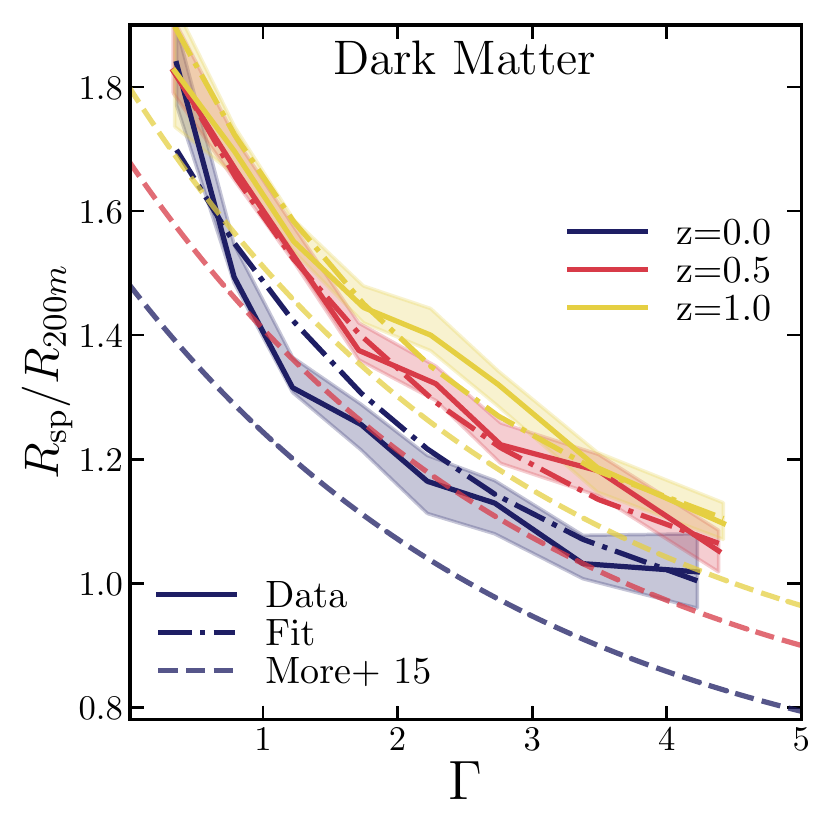}
    \vspace{-.5cm}
    \caption{The location of the splashback radius as a function of accretion rate $\Gamma$ for the dark matter in the hydrodynamic simulation.  The measurements from TNG300-1 are shown in the solid lines with the 16th and 84th percentiles of our bootstrap sample shown by the shaded bands.  We fit a function following the form of Equation \ref{eq:gamma_fit}, shown with the dot-dashed lines.  The original model from \citet{More2015} is shown with the dashed lines.  Our measurements are consistently higher than the \citet{More2015} model, which results in a different fit.  Although the modified fit is in better agreement with our results, there is still a difference between our measurements and the model in the shape of the lines and the spacing between redshifts.}
    \label{fig:gamma_fit}
\end{figure}

\begin{figure}
    \centering
    \includegraphics[width=\linewidth]{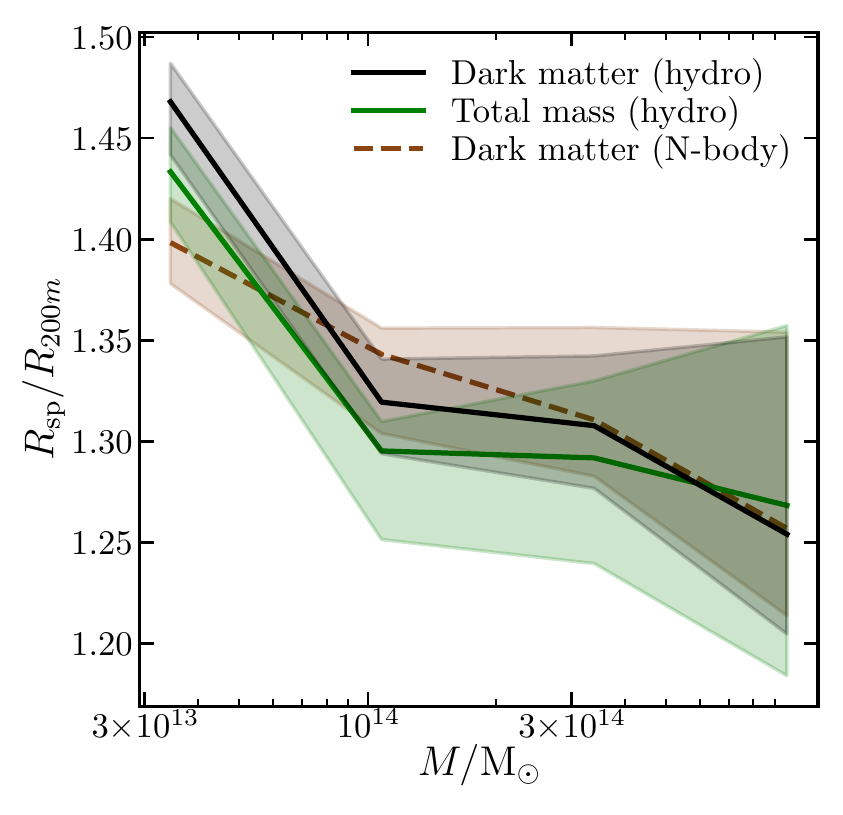}
    \vspace{-.3cm}
    \caption{Location of the splashback radius as a function of halo mass for the dark matter and total mass in a hydrodynamic simulation (solid lines) and the dark matter mass in an \textit{N}-body simulation (dashed line) at $z=0$.  We stack the density profiles for haloes with $\log_{10}\left(M_{200\rm{m}}/{\rm M}_{\odot}\right)$ in ${13}-{13.5}$, ${13.5}-{14}$, ${14}-{14.5}$ and ${14.5}-{15}$ and compute the splashback radius of the median profile, shown by the solid and dashed lines.  To estimate error, we use a bootstrap method and show the 16th and 84th percentiles as the shaded band around each line.  We do not find a significant difference between the hydrodynamic and \textit{N}-body simulations for $R_{\rm sp}/R_{\rm200m}$ as a function of halo mass.}
    \label{fig:dm_mass_evolution}
\end{figure}

\begin{figure*}
    \centering
    \includegraphics[width=\linewidth]{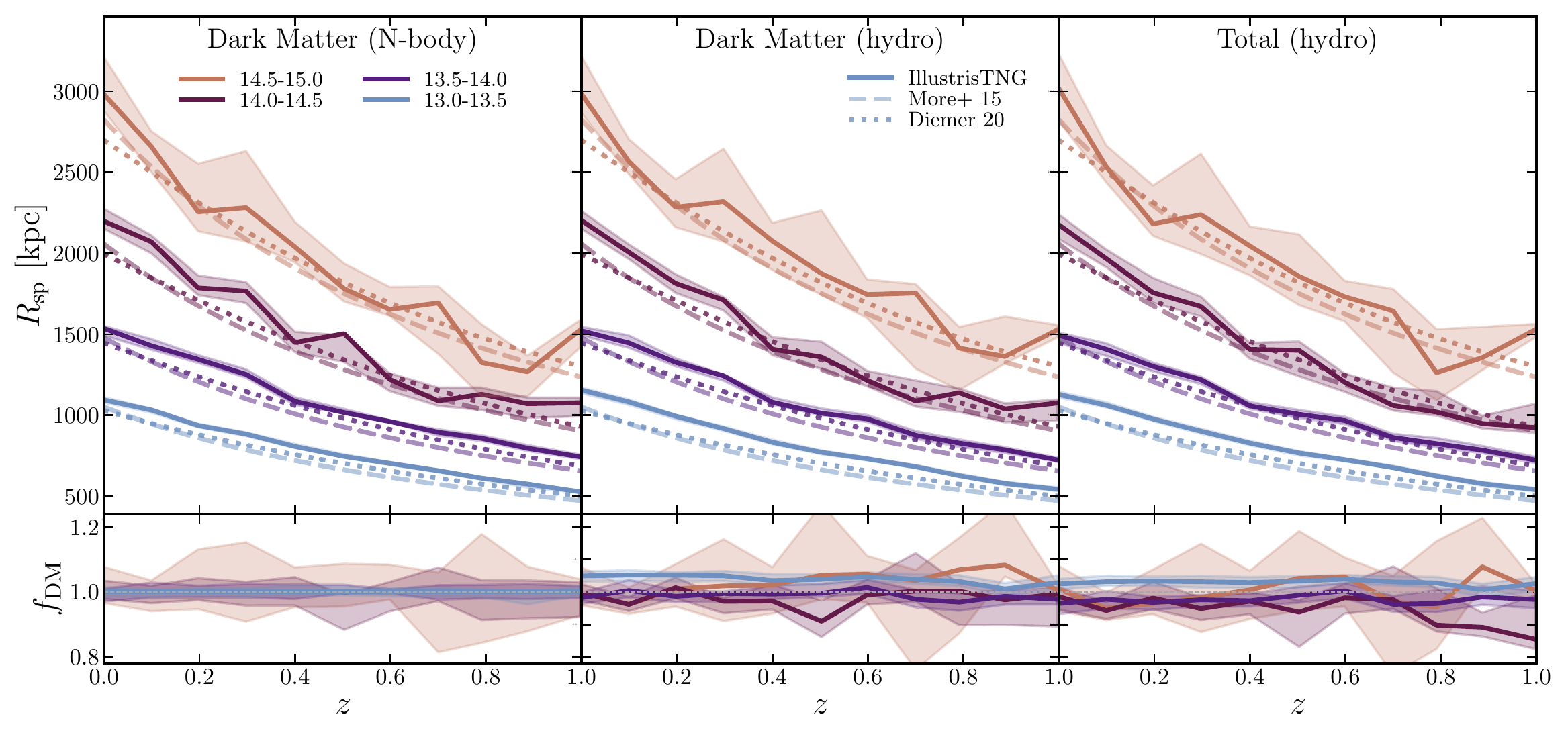}
    \vspace{-.3cm}
    \caption{The top panels show the splashback radius as a function of redshift for four halo mass ranges.  Here we compare results from the dark matter mass density in the \textit{N}-body simulation TNG300-1-DM (left) to the dark matter mass (middle) and total mass (right) density profiles in the hydrodynamic simulation TNG300-1 .  We stack the density profiles for haloes in four mass bins between $10^{13}\rm{M}_{\odot}$ and $10^{15}\rm{M}_{\odot}$ and compute the splashback radius of the median profile, shown by the solid lines.  Each colour corresponds to a mass bin labeled with the $\log_{10}$ of its bounds.  To estimate error, we use a bootstrap method and show the 16th and 84th percentiles as the shaded band around each line.  We compare to two analytic models, found in \citet{More2015} and \citet{Diemer2020a}, using the middle of each mass range ($\log(M/{\rm M}_{\odot}) = 13.25, 13.75, 14.25$ and $14.75$) at each redshift.  These models predict $R_{\rm sp}$ for dark matter haloes, but we show them in each panel for reference.  The bottom panels show the fractional difference with the \textit{N}-body simulations.  The lowest mass bin is slightly higher in the hydrodynamic than in the \textit{N}-body simulation, but there is overall agreement between the two across all redshifts.}
    \label{fig:dm_redshift_evolution}
\end{figure*}

Previous work, e.g. \citet{Diemer2014, Adhikari2014, More2015, Mansfield2017, Diemer2017b}, has found a strong correlation between the splashback radius and halo accretion rate, $\Gamma$.
The mass dependence of $R_{\rm sp}$, discussed in Section \ref{sec:results_mass}, is typically attributed to the tendency for larger haloes to have higher accretion rates \citep[e.g.][]{Diemer2014}.
In this section, we examine the relationship of $R_{\rm sp}$ and $\Gamma$ to properly compare to these models.

Several models have been proposed in these past studies for predicting the location of the splashback radius based on the accretion rate of haloes.
In general, $R_{\rm sp}/R_{\rm200m}$ decreases for higher accretion rates.
For a high accretion rate, the potential deepens more quickly, causing splashback to occur at a smaller radius \citep{Adhikari2014} as discussed in Section \ref{sec:results_redshift}.
At a constant accretion rate, $R_{\rm sp}/R_{\rm200m}$ decreases with $z$, which is predicted to be dependence on $\Omega_{\rm m}$.
At higher redshifts, $R_{\rm200m}$ will be smaller for haloes of similar masses, so the ratio $R_{\rm sp}/R_{\rm200m}$ will increase.
Increasing $R_{\rm sp}/R_{\rm200m}$ with redshift for a constant accretion rate agrees with the $\Omega_{\rm m}$ dependence discussed in \citet{Diemer2017b}.
For comparison with previous work, which related $R_{\rm sp}/R_{\rm200m}$ to $\Gamma$, we plot the ratio of $R_{\rm sp}$ and $R_{\rm200m}$ rather than $R_{\rm sp}$.

At low accretion rate ($\Gamma<1)$, a second caustic appears at a smaller radius than $R_{\rm sp}$ in the dark matter profiles, visible in Figure \ref{fig:caustic}.
This has also been observed in previous work, and it is likely due to the more distinct streams of infalling material for lower accretion rates \citep[e.g.][]{Adhikari2014,Deason2020a}.
There is some evidence of this in our galaxy profiles, although the large spacing of linear bins at smaller radii and increased noise compared to the dark matter profiles makes it difficult to identify, and it does not appear in our gas profiles.
For suitable fit parameters, the minimum value of the lograthmic derivative is dominated by the splashback feature even at low accretion rates (see Figure \ref{fig:caustic}).
Since we are interested in the properties of the splashback feature, we do not attempt to include this second caustic in our fits.
We have checked that our fitting method consistently finds the appropriate minimum.

Like the mass dependence, we find that our results follow similar trends as previous work but that $R_{\rm sp}$, as calculated using the point of steepest slope, is slightly higher compared to previous work.
In Figure \ref{fig:gamma_components}, we show the location of the splashback radius, $R_{\rm sp}/R_{\rm200m}$, for the dark matter mass, gas mass and galaxy number density profiles as a function of halo accretion rate between $z=0$ and $1$.
We compare to the \citet{More2015} and \citet{Diemer2020a} models with the dashed and dotted lines, respectively.
As a function of redshift $z$ and accretion rate $\Gamma$, \citet{More2015} find
\begin{equation}
    \indent\frac{R_{\rm sp}}{R_{\rm200m}} = 0.54\left[1+0.53\Omega_{\rm m}(z)\right] \left(1+1.36e^{-\Gamma/3.04}\right)\:.
    \label{eq:more15}
\end{equation}
\citet{Diemer2017b} and \citet{Diemer2020a} propose a similar form for $R_{\rm sp}$:
\begin{equation}
   \indent R_{\rm sp} = A + Be^{-\Gamma/C}\:,
\end{equation}
where $A, B, C$ contain a dependence on $\Omega_{\rm m}$ and peak height.
These parameters also contain the dependence on the percentile of first apocentres enclosed by $R_{\rm sp}$, which we set to 0.75 for our comparisons.
As in Figure \ref{fig:redshift_evolution_components}, we show these models in all three panels for reference, though they were developed from dark matter only simulations.

The \citet{Diemer2020a} model converges across redshifts at low $\Gamma$ more than the \citet{More2015} model.
Despite following a method that more closely resembles that of \citet{More2015}, we also find that $R_{\rm sp}/R_{\rm200m}$ converges for low $\Gamma$ in both the dark matter and galaxy profiles.
However, given the noise in our samples, it is difficult to determine the significance of this observation.

The gas profiles again yield lower results than the dark matter profiles and the deviation depends on accretion rate more than on redshift.
Across all redshifts, the lowest accretion rate bin has an $R_{\rm sp}/R_{\rm200m}$ value that is slightly lower for the gas profiles, but there is significant variance across the redshifts.
For $\Gamma$ between $0.5-3$, $R_{\rm sp}/R_{\rm200m}$ is $\sim10\%$ lower for the gas.
For $\Gamma$ between 3 and 4, $R_{\rm sp}/R_{\rm200m}$ is $\sim5\%$ lower for the gas, and we do not find a significant difference in the highest accretion rate bin between the gas and dark matter profiles.

\citet{Xhakaj2020} found that subhalo profiles produce a significantly lower $R_{\rm sp}$ than dark matter profiles for all accretion rates.
While on average, $R_{\rm sp}$ for the galaxy profiles is slightly lower than the dark matter profiles, we do not find a significant difference.
As noted in Section \ref{sec:results_mass}, \citet{Xhakaj2020} binned haloes by accretion rate over a narrow mass range near $10^{14}\rm{M}_{\odot}$ and used subhaloes near $10^{12}\rm{M}_{\odot}$.
It is possible that this difference is washed out in our profiles due to wider mass ranges and the statistical noise of our sample.  Our subhalo sample is also significantly less massive, with masses down to $10^{9}\:\rm{M}_{\odot}$.  The effect of subhalo populations on $R_{\rm sp}$ and effects of the fitting method on subhalo profiles will be investigated in future work.

Since our results are consistently higher than the predictions from previous work, we fit a function based on the \citet{More2015} model to test if our results can be well described by these models.  We fit a modified version of Equation \ref{eq:more15}:
\begin{equation}
    \label{eq:gamma_fit}
    \indent\frac{R_{\rm sp}}{R_{\rm200m}} = A[1+B\Omega_{\rm m}(z)]\left(1+Ce^{-\Gamma/D}\right)
\end{equation}
and obtain $A=0.80, B=0.26, C=1.14,$ and $D=1.25$ for the dark matter profiles in TNG300-1, which are significanly different from the values in Equation \ref{eq:more15}.  We fit all redshifts simultaneously, so each profile has two independent variables, $\Gamma$ and $z$.  Figure \ref{fig:gamma_fit} shows this fit along with the original \citet{More2015} model and our measurements.

The modified \citet{More2015} model is in better agreement with our results, although there is still a discrepancy in the shape of the curves and spacing between redshifts.  Given the noise in our profiles, it is difficult to determine the significance of this discrepancy.  We do not attempt to fit a modified \citet{Diemer2020a} model since this model depends on parameters obtained from tracing the trajectories of particles.  We leave a more detailed exploration of a functional form that better describes our results to future work.

\begin{figure*}
    \centering
    \includegraphics[width=\linewidth]{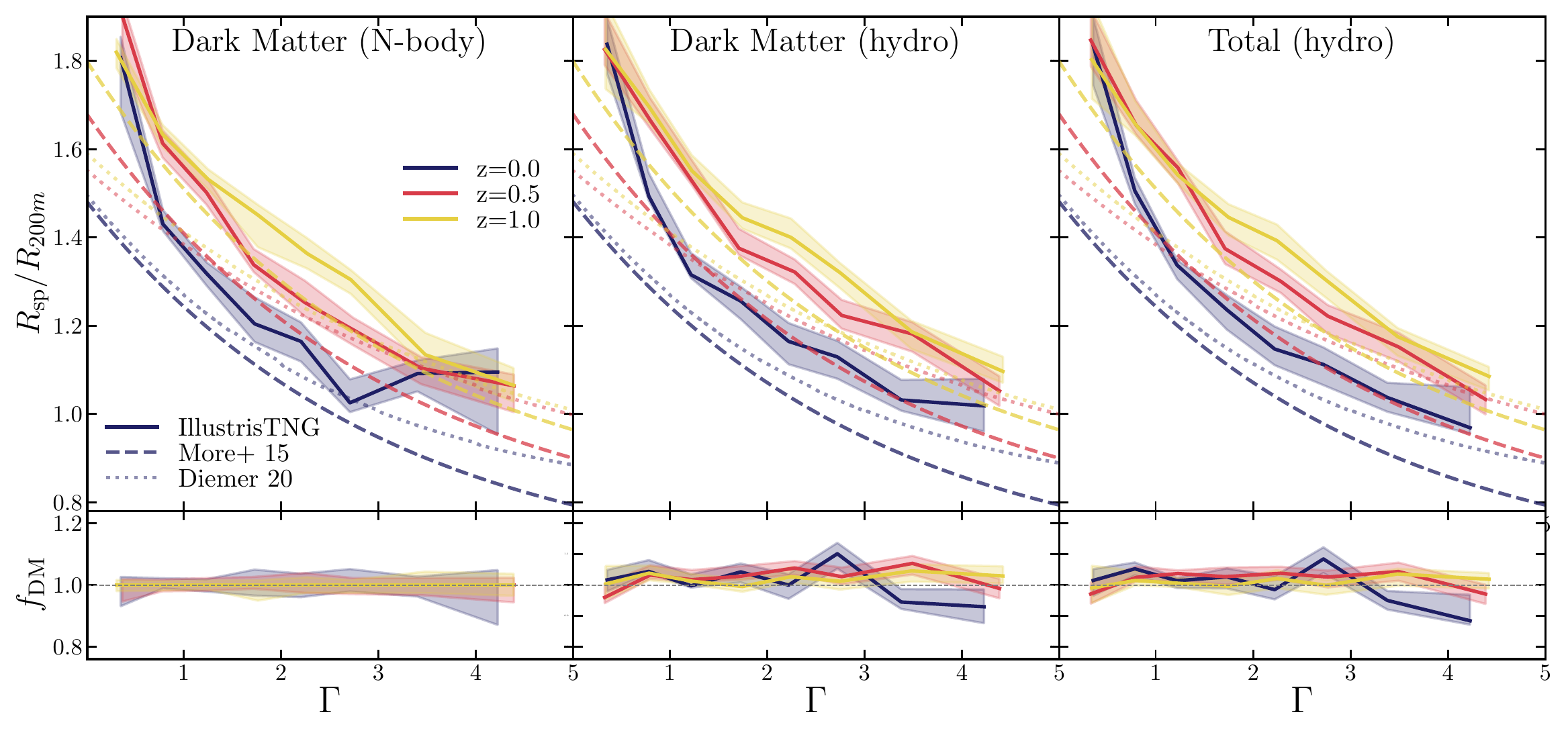}
    \vspace{-.3cm}
    \caption{The top panels show the location of the splashback radius as a function of accretion rate $\Gamma$ for the dark matter mass in TNG300-1-DM (left) and the dark matter (middle) and total mass (right) in TNG300-1.  The colours show several redshifts, and we compare to the models described in \citet{More2015} and \citet{Diemer2020a} shown in the dashed and dotted lines repsectively.  The location of the splashback radius decreases with accretion rate and increases with redshift.  The haloes are stacked based on accretion rates in eight bins between $0$ and $5$.  We compute the splashback radius of the median profile for each accretion range.  To compare to the analytic models, we calculate a value for $R_{\rm sp}/R_{\rm 200m}$ based on the median $R_{\rm200m}$ value for each accretion rate range.  These models predict $R_{\rm sp}$ for dark matter haloes, but we show them in each panel for reference.  The bottom panels show the fractional differences between $R_{\rm sp}/R_{\rm 200m}$ for TNG300-1-DM and the dark matter and total mass density profiles in TNG300-1.}
    \label{fig:gamma_sims}
\end{figure*}

\subsection{Comparison between \textit{N}-body and hydrodynamic simulations}
\label{sec:results_dm}

The majority of previous work has been done using dark matter only simulations.
We explore the impact of baryonic physics on the splashback feature by comparing the magnetohydrodynamic TNG300 runs to the \textit{N}-body, dark matter only TNG300-DM runs.
We use both the dark matter and total mass (dark matter, gas, and stars) in the TNG300 simulations to compare to the dark matter mass in the TNG300-DM simulations.
In this section, we focus on the highest resolution (level 1) runs.

We follow the same procedure for this comparison as was used in the previous Section.
The haloes are split into four mass or eight accretion rate bins and we measure $R_{\rm sp}$ for the median stacked profile at redshifts ranging between 0 and 1.

We show comparisons for $R_{\rm sp}/R_{\rm200m}$ as a function of mass between in Figure \ref{fig:dm_mass_evolution}.
Similarly to the total mass and dark matter profiles in the hydrodynamic simulation, $R_{\rm sp}/R_{\rm200m}$ decreases with increasing halo mass, and we do not find a significant difference with the introduction of the baryonic component and galaxy formation processes.

$R_{\rm sp}$ for each mass bin is also shown as a function of redshift in Figure \ref{fig:dm_redshift_evolution}, with the \citet{More2015} and \citet{Diemer2020a} shown in the dashed and dotted lines.
We do not find a significant difference in the results between either the dark matter or total mass in TNG300-1 and the dark matter mass in TNG300-1-DM.
The largest difference occurs in the smallest smallest mass bin, where the deviation between TNG300-1 and TNG300-1-DM is less than $5\%$.
Given the inaccuracies inherent in our simulation data and profile fits, it is not clear that the small differences we measure are statistically significant.

We show $R_{\rm sp}/R_{\rm200m}$ again as a function of accretion rate for several redshifts in Figure \ref{fig:gamma_sims}.
We find no significant difference between $R_{\rm sp}$ in the dark matter and total mass profiles in the hydrodynamic run and the dark matter profiles in the \textit{N}-body run for all accretion rates and redshifts.
As in Figure \ref{fig:dm_redshift_evolution}, we compare to the \citet{More2015} and \citet{Diemer2020a} models shown in the dashed and dotted lines respectively.

Overall, the results from the TNG300-1 run agree well with the TNG300-1-DM run.

\section{Conclusions}
\label{sec:conclusions}
We have explored the splashback radius, $R_{\rm sp}$, in the TNG300 volume of the IllustrisTNG simulations and compared to its \textit{N}-body dark matter only counterpart TNG300-DM.
The methods developed in dark matter only simulations work well in hydrodynamic simulations, and the addition of baryonic physics has minimal impact on our work.
Defining $R_{\rm sp}$ as the minimum of the derivative of the spherically averaged density profile, we computed $R_{\rm sp}$ for haloes with $10^{13}\rm{M}_{\odot}<M_{\rm200m}<10^{15}\rm{M}_{\odot}$.
Our conclusions are summarised as follows:
\begin{itemize}
    \item The identification of the point of steepest slope in the density profile relies on choices in describing the density profile, and the method used to fit the density profile systematically influences the results.  Methods that better fit the sharp decrease in the derivative of the profile have a bias towards larger $R_{\rm sp}$ (Figure \ref{fig:fit_comp}).  Our results are most robust when we fit the derivative of the density profile directly to find its minimum.  We leave eight parameters in the fitting function free to vary.  Our calculated $\chi^2$ value for this method remains low relative to other fitting methods (e.g. fitting the density profile and then differentiating) for our range of halo mass, halo accretion rate and redshift (Figure \ref{fig:chi2}).  Our calculations yield a $R_{\rm sp}$ value that is generally higher than other methods  by $\sim5\%$ (Figure \ref{fig:avg_fractional_fit}).
    \item We calculated the density profiles of the dark matter, gas and galaxies in haloes and fit each density profile (Figure \ref{fig:stacked_density_examples}).  We stacked haloes in four logarithmically spaced mass bins between $10^{13}\rm{M}_{\odot}-10^{15}\rm{M}_{\odot}$ and found that $R_{\rm sp}/R_{\rm200m}$ decreases with $M_{\rm200m}$ (Figure \ref{fig:mass_evolution}).
    This trend is most evident in the dark matter and gas profiles.
    
    \item The steepest slope computed from the gas profile is $\sim10-20\%$ lower than computed using the dark matter profile.
    The difference is larger for haloes with larger masses (Figure \ref{fig:mass_evolution}).
    This is not surprising given that the gas dynamics is governed by shocks rather than collisionless orbital dynamics.
    
    \item $R_{\rm sp}$ computed from galaxy number density profiles is similar to the value from dark matter profiles for haloes with $M>10^{13.5}\rm{M}_{\odot}$ (also see Figure \ref{fig:mass_evolution}).
    For haloes with $10^{13}\rm{M}_{\odot}<M<10^{13.5}\rm{M}_{\odot}$, $R_{\rm sp}$ is $\sim12\%$ lower in the galaxy profiles than in the dark matter profiles.
    This is likely due to dynamical friction as galaxies fall into less massive haloes.
    
    \item We computed $R_{\rm sp}$ for redshifts $0-1$ and found that $R_{\rm sp}$ decreases with with redshift for haloes with similar $M_{\rm200m}$ (Figure \ref{fig:redshift_evolution_components}).
    The dark matter, gas and galaxy profiles yield similar dependence on redshift.
    The differences computed in $R_{\rm sp}$ for a given mass are consistent across this redshift range.
    
    \item We also stacked our sample of haloes in eight accretion rate bins between $0$ and $5$.  At lower accretion rate ($\Gamma<1$), a second, smaller caustic appears at a smaller radius than $R_{\rm sp}$, but this does not affect our results (Figure \ref{fig:caustic}).  $R_{\rm sp}/R_{\rm200m}$ decreases with accretion rate (Figure \ref{fig:gamma_components}).
    This is consistent with previous work, although our values are consistently higher.
    We fit a modified version of the $R_{\rm sp}$ model proposed in \citet{More2015} to our data that produces better agreement, but the shapes of our curves still differ from the \citet{More2015} model (Figure \ref{fig:gamma_fit}).
    A decrease with accretion rate is expected because haloes with a higher accretion rate will have a larger increase in potential after one orbit, causing the first apocentre to occur at a smaller radius.
    
    \item The dark matter and total mass density profiles in the hydrodynamic simulation TNG300 yield results similar to the dark matter profiles in the dark matter only TNG300-DM simulation, with differences less than $5\%$ (Figures \ref{fig:dm_mass_evolution}, \ref{fig:dm_redshift_evolution} and \ref{fig:gamma_sims}).
    We therefore conclude that the addition of baryonic physics does not significantly influence the dark matter dyanamics near $R_{\rm sp}$.
\end{itemize}

Comparison with previous work is complicated by the differences in methods used to identify $R_{\rm sp}$ and by systematic differences caused by the method used to fit a given profile.
However, we demonstrate qualitative agreement with past work, which indicates that $R_{\rm sp}$ could be a reliable measure of halo size.
Given the minimal impact due to baryonic physics on the value of $R_{\rm sp}$, results from dark matter only simulations should produce reliable predictions in more physical contexts.
Additionally, a splashback-\textit{like} feature is evident in the observable components of haloes, gas and galaxies, although they differ somewhat in underlying physics.
This is a promising indication that $R_{\rm sp}$ could be inferred from optical and X-ray observations.
Initial studies have been done in existing surveys, e.g. SDSS, and this holds the potential to be more widely used in observations from eROSITA, the Nancy Grace Roman Space Telescope and the James Webb Space Telescope.
Future work will be needed to better understand any biases introduced by observational constraints.
An extension of the work presented here would be to study the effect of galaxy definition and introduce magnitude limits on subhaloes included in the galaxy sample.

\section*{Acknowledgements}
We thank the anonymous referee for providing comments that improved this manuscript.
This analysis was performed using the MIT/Harvard computing facilities supported by FAS and MKI.
MV acknowledges support through NASA ATP grants 16-ATP16-0167, 19-ATP19-0019, 19-ATP19-0020, 19-ATP19-0167, and NSF grants AST-1814053, AST-1814259,  AST-1909831 and AST-2007355.

\section*{Data Availability}
The data used in this work can be accessed at \href{www.tng-project.org}{www.tng-project.org}.

\begin{table*}
\begin{tabular}{c|ccccccccccc}
      Simulation & \multicolumn{11}{c}{$z$} \\
      & 0.0 & 0.1 & 0.2 & 0.3 & 0.4 & 0.5 & 0.6 & 0.7 & 0.8 & 0.9 & 1.0 \\
     \hline
     TNG300-1 & 1401 & 1447 & 1496 & 1558 & 1578 & 1568 & 1568 & 1546 & 1506 & 1459 & 1393 \\
     TNG300-2 & 1412 & 1453 & 1505 & 1545 & 1584 & 1578 & 1563 & 1536 & 1499 & 1468 & 1387 \\
     TNG300-3 & 1431 & 1465 & 1526 & 1574 & 1570 & 1563 & 1588 & 1559 & 1528 & 1481 & 1405 \\
\end{tabular}
\caption{The number of haloes included in our sample in each resolution level (1, 2 and 3 for high, medium and low resolution) of the full hydrodynamic (TNG300) runs.  We select haloes for redshifts $0\leq z\leq 1$ in increments of 0.1.  This number is the number of Friends-of-Friends groups with $M_{\rm200m}>10^{13}\rm{M}_{\odot}$, and haloes within $10\times R_{\rm200m}$ of a larger halo are not counted.  See Table \ref{table:res_sample_size} for the number of haloes in each mass and accretion rate range we examined.}
\label{table:res_tot_sample_size}
\end{table*}

\bibliographystyle{mnras}
\bibliography{bibliography}

\begin{thebibliography}{}
\makeatletter
\relax
\def\mn@urlcharsother{\let\do\@makeother \do\$\do\&\do\#\do\^\do\_\do\%\do\~}
\def\mn@doi{\begingroup\mn@urlcharsother \@ifnextchar [ {\mn@doi@}
  {\mn@doi@[]}}
\def\mn@doi@[#1]#2{\def\@tempa{#1}\ifx\@tempa\@empty \href
  {http://dx.doi.org/#2} {doi:#2}\else \href {http://dx.doi.org/#2} {#1}\fi
  \endgroup}
\def\mn@eprint#1#2{\mn@eprint@#1:#2::\@nil}
\def\mn@eprint@arXiv#1{\href {http://arxiv.org/abs/#1} {{\tt arXiv:#1}}}
\def\mn@eprint@dblp#1{\href {http://dblp.uni-trier.de/rec/bibtex/#1.xml}
  {dblp:#1}}
\def\mn@eprint@#1:#2:#3:#4\@nil{\def\@tempa {#1}\def\@tempb {#2}\def\@tempc
  {#3}\ifx \@tempc \@empty \let \@tempc \@tempb \let \@tempb \@tempa \fi \ifx
  \@tempb \@empty \def\@tempb {arXiv}\fi \@ifundefined
  {mn@eprint@\@tempb}{\@tempb:\@tempc}{\expandafter \expandafter \csname
  mn@eprint@\@tempb\endcsname \expandafter{\@tempc}}}

\bibitem[\protect\citeauthoryear{{Adhikari}, {Dalal}  \&
  {Chamberlain}}{{Adhikari} et~al.}{2014}]{Adhikari2014}
{Adhikari} S.,  {Dalal} N.,   {Chamberlain} R.~T.,  2014, \mn@doi [\jcap]
  {10.1088/1475-7516/2014/11/019}, \href
  {https://ui.adsabs.harvard.edu/abs/2014JCAP...11..019A} {2014, 019}

\bibitem[\protect\citeauthoryear{{Adhikari}, {Dalal}  \& {Clampitt}}{{Adhikari}
  et~al.}{2016}]{Adhikari2016}
{Adhikari} S.,  {Dalal} N.,   {Clampitt} J.,  2016, \mn@doi [\jcap]
  {10.1088/1475-7516/2016/07/022}, \href
  {https://ui.adsabs.harvard.edu/abs/2016JCAP...07..022A} {2016, 022}

\bibitem[\protect\citeauthoryear{{Adhikari}, {Sakstein}, {Jain}, {Dalal}  \&
  {Li}}{{Adhikari} et~al.}{2018}]{Adhikari2018}
{Adhikari} S.,  {Sakstein} J.,  {Jain} B.,  {Dalal} N.,   {Li} B.,  2018,
  \mn@doi [\jcap] {10.1088/1475-7516/2018/11/033}, \href
  {https://ui.adsabs.harvard.edu/abs/2018JCAP...11..033A} {2018, 033}

\bibitem[\protect\citeauthoryear{{Adhikari} et~al.,}{{Adhikari}
  et~al.}{2020}]{Adhikari2020}
{Adhikari} S.,  et~al., 2020, arXiv e-prints, \href
  {https://ui.adsabs.harvard.edu/abs/2020arXiv200811663A} {p. arXiv:2008.11663}

\bibitem[\protect\citeauthoryear{{Allen}, {Evrard}  \& {Mantz}}{{Allen}
  et~al.}{2011}]{Allen2011}
{Allen} S.~W.,  {Evrard} A.~E.,   {Mantz} A.~B.,  2011, \mn@doi [Annual Review
  of Astronomy and Astrophysics] {10.1146/annurev-astro-081710-102514}, \href
  {https://ui.adsabs.harvard.edu/abs/2011ARA\%26A..49..409A/abstract} {49, 409}

\bibitem[\protect\citeauthoryear{{Aung}, {Nagai}  \& {Lau}}{{Aung}
  et~al.}{2020}]{Aung2020}
{Aung} H.,  {Nagai} D.,   {Lau} E.~T.,  2020, arXiv e-prints, \href
  {https://ui.adsabs.harvard.edu/abs/2020arXiv201200977A} {p. arXiv:2012.00977}

\bibitem[\protect\citeauthoryear{{Banerjee}, {Adhikari}, {Dalal}, {More}  \&
  {Kravtsov}}{{Banerjee} et~al.}{2020}]{Banerjee2020}
{Banerjee} A.,  {Adhikari} S.,  {Dalal} N.,  {More} S.,   {Kravtsov} A.,  2020,
  \mn@doi [\jcap] {10.1088/1475-7516/2020/02/024}, \href
  {https://ui.adsabs.harvard.edu/abs/2020JCAP...02..024B} {2020, 024}

\bibitem[\protect\citeauthoryear{{Barnes}, {Kay}, {Henson}, {McCarthy},
  {Schaye}  \& {Jenkins}}{{Barnes} et~al.}{2017a}]{Barnes2017a}
{Barnes} D.~J.,  {Kay} S.~T.,  {Henson} M.~A.,  {McCarthy} I.~G.,  {Schaye} J.,
    {Jenkins} A.,  2017a, \mn@doi [\mnras] {10.1093/mnras/stw2722}, \href
  {https://ui.adsabs.harvard.edu/abs/2017MNRAS.465..213B} {465, 213}

\bibitem[\protect\citeauthoryear{{Barnes} et~al.,}{{Barnes}
  et~al.}{2017b}]{Barnes2017b}
{Barnes} D.~J.,  et~al., 2017b, \mn@doi [\mnras] {10.1093/mnras/stx1647}, \href
  {https://ui.adsabs.harvard.edu/abs/2017MNRAS.471.1088B} {471, 1088}

\bibitem[\protect\citeauthoryear{{Barnes} et~al.,}{{Barnes}
  et~al.}{2018}]{Barnes2018a}
{Barnes} D.~J.,  et~al., 2018, \mn@doi [\mnras] {10.1093/mnras/sty2078}, \href
  {https://ui.adsabs.harvard.edu/abs/2018MNRAS.481.1809B/abstract} {481, 1809}

\bibitem[\protect\citeauthoryear{{Barnes}, {Vogelsberger}, {Pearce}, {Pop},
  {Kannan}, {Cao}, {Kay}  \& {Hernquist}}{{Barnes} et~al.}{2021}]{Barnes2020a}
{Barnes} D.~J.,  {Vogelsberger} M.,  {Pearce} F.~A.,  {Pop} A.-R.,  {Kannan}
  R.,  {Cao} K.,  {Kay} S.~T.,   {Hernquist} L.,  2021, \mn@doi [\mnras]
  {10.1093/mnras/stab1276}, \href
  {https://ui.adsabs.harvard.edu/abs/2021MNRAS.tmp.1271B} {}

\bibitem[\protect\citeauthoryear{{Baxter} et~al.,}{{Baxter}
  et~al.}{2015}]{Baxter2015}
{Baxter} E.~J.,  et~al., 2015, \mn@doi [\apj] {10.1088/0004-637X/806/2/247},
  \href {https://ui.adsabs.harvard.edu/abs/2015ApJ...806..247B} {806, 247}

\bibitem[\protect\citeauthoryear{{Baxter} et~al.,}{{Baxter}
  et~al.}{2017}]{Baxter2017}
{Baxter} E.,  et~al., 2017, \mn@doi [\apj] {10.3847/1538-4357/aa6ff0}, \href
  {https://ui.adsabs.harvard.edu/abs/2017ApJ...841...18B} {841, 18}

\bibitem[\protect\citeauthoryear{{Bertschinger}}{{Bertschinger}}{1985}]{Bertschinger1985}
{Bertschinger} E.,  1985, \mn@doi [\apjs] {10.1086/191028}, \href
  {https://ui.adsabs.harvard.edu/abs/1985ApJS...58...39B/abstract} {58, 39}

\bibitem[\protect\citeauthoryear{{Biffi}, {Sembolini}, {De Petris},
  {Valdarnini}, {Yepes}  \& {Gottl{\"o}ber}}{{Biffi} et~al.}{2014}]{Biffi2014}
{Biffi} V.,  {Sembolini} F.,  {De Petris} M.,  {Valdarnini} R.,  {Yepes} G.,
  {Gottl{\"o}ber} S.,  2014, \mn@doi [\mnras] {10.1093/mnras/stu018}, \href
  {https://ui.adsabs.harvard.edu/abs/2014MNRAS.439..588B} {439, 588}

\bibitem[\protect\citeauthoryear{{Blumenthal}, {Faber}, {Primack}  \&
  {Rees}}{{Blumenthal} et~al.}{1984}]{Blumenthal1984}
{Blumenthal} G.~R.,  {Faber} S.~M.,  {Primack} J.~R.,   {Rees} M.~J.,  1984,
  \mn@doi [\nat] {10.1038/311517a0}, \href
  {https://ui.adsabs.harvard.edu/abs/1984Natur.311..517B} {311, 517}

\bibitem[\protect\citeauthoryear{{Bocquet} et~al.,}{{Bocquet}
  et~al.}{2019}]{Bocquet2019}
{Bocquet} S.,  et~al., 2019, \mn@doi [\apj] {10.3847/1538-4357/ab1f10}, \href
  {https://ui.adsabs.harvard.edu/abs/2019ApJ...878...55B} {878, 55}

\bibitem[\protect\citeauthoryear{{Bond}, {Kofman}  \& {Pogosyan}}{{Bond}
  et~al.}{1996}]{Bond1996b}
{Bond} J.~R.,  {Kofman} L.,   {Pogosyan} D.,  1996, \mn@doi [\nat]
  {10.1038/380603a0}, \href
  {https://ui.adsabs.harvard.edu/abs/1996Natur.380..603B} {380, 603}

\bibitem[\protect\citeauthoryear{{Bryan} \& {Norman}}{{Bryan} \&
  {Norman}}{1998}]{Bryan1998}
{Bryan} G.~L.,  {Norman} M.~L.,  1998, \mn@doi [\apj] {10.1086/305262}, \href
  {https://ui.adsabs.harvard.edu/abs/1998ApJ...495...80B} {495, 80}

\bibitem[\protect\citeauthoryear{{Chang} et~al.,}{{Chang}
  et~al.}{2018}]{Chang2018}
{Chang} C.,  et~al., 2018, \mn@doi [\apj] {10.3847/1538-4357/aad5e7}, \href
  {https://ui.adsabs.harvard.edu/abs/2018ApJ...864...83C/abstract} {864, 18}

\bibitem[\protect\citeauthoryear{{Contigiani}, {Vardanyan}  \&
  {Silvestri}}{{Contigiani} et~al.}{2019}]{Contigiani2019}
{Contigiani} O.,  {Vardanyan} V.,   {Silvestri} A.,  2019, \mn@doi [\prd]
  {10.1103/PhysRevD.99.064030}, \href
  {https://ui.adsabs.harvard.edu/abs/2019PhRvD..99f4030C} {99, 064030}

\bibitem[\protect\citeauthoryear{{Cuesta}, {Prada}, {Klypin}  \&
  {Moles}}{{Cuesta} et~al.}{2008}]{Cuesta2008}
{Cuesta} A.~J.,  {Prada} F.,  {Klypin} A.,   {Moles} M.,  2008, \mn@doi
  [\mnras] {10.1111/j.1365-2966.2008.13590.x}, \href
  {https://ui.adsabs.harvard.edu/abs/2008MNRAS.389..385C} {389, 385}

\bibitem[\protect\citeauthoryear{{Davis}, {Efstathiou}, {Frenk}  \&
  {White}}{{Davis} et~al.}{1985}]{Davis1985}
{Davis} M.,  {Efstathiou} G.,  {Frenk} C.~S.,   {White} S.~D.~M.,  1985,
  \mn@doi [\apj] {10.1086/163168}, \href
  {https://ui.adsabs.harvard.edu/abs/1985ApJ...292..371D/abstract} {292, 371}

\bibitem[\protect\citeauthoryear{{Deason}, {Fattahi}, {Frenk}, {Grand}, {Oman},
  {Garrison-Kimmel}, {Simpson}  \& {Navarro}}{{Deason}
  et~al.}{2020}]{Deason2020a}
{Deason} A.~J.,  {Fattahi} A.,  {Frenk} C.~S.,  {Grand} R. J.~J.,  {Oman}
  K.~A.,  {Garrison-Kimmel} S.,  {Simpson} C.~M.,   {Navarro} J.~F.,  2020,
  \mn@doi [\mnras] {10.1093/mnras/staa1711}, \href
  {https://ui.adsabs.harvard.edu/abs/2020MNRAS.496.3929D} {496, 3929}

\bibitem[\protect\citeauthoryear{{Deason} et~al.,}{{Deason}
  et~al.}{2021}]{Deason2020b}
{Deason} A.~J.,  et~al., 2021, \mn@doi [\mnras] {10.1093/mnras/staa3590}, \href
  {https://ui.adsabs.harvard.edu/abs/2021MNRAS.500.4181D} {500, 4181}

\bibitem[\protect\citeauthoryear{{Di Matteo}, {Springel}  \& {Hernquist}}{{Di
  Matteo} et~al.}{2005}]{DiMatteo2005}
{Di Matteo} T.,  {Springel} V.,   {Hernquist} L.,  2005, \mn@doi [\nat]
  {10.1038/nature03335}, \href
  {http://adsabs.harvard.edu/abs/2005Natur.433..604D} {433, 604}

\bibitem[\protect\citeauthoryear{{Diemer}}{{Diemer}}{2017}]{Diemer2017a}
{Diemer} B.,  2017, \mn@doi [\apjs] {10.3847/1538-4365/aa799c}, \href
  {https://ui.adsabs.harvard.edu/abs/2017ApJS..231....5D} {231, 5}

\bibitem[\protect\citeauthoryear{{Diemer}}{{Diemer}}{2018}]{Diemer2018}
{Diemer} B.,  2018, \mn@doi [\apj] {10.3847/1538-4365/aaee8c}, \href
  {https://ui.adsabs.harvard.edu/abs/2018ApJS..239...35D/abstract} {239, 13}

\bibitem[\protect\citeauthoryear{{Diemer}}{{Diemer}}{2020}]{Diemer2020a}
{Diemer} B.,  2020, \mn@doi [\apjs] {10.3847/1538-4365/abbf51}, \href
  {https://ui.adsabs.harvard.edu/abs/2020ApJS..251...17D} {251, 17}

\bibitem[\protect\citeauthoryear{{Diemer} \& {Kravtsov}}{{Diemer} \&
  {Kravtsov}}{2014}]{Diemer2014}
{Diemer} B.,  {Kravtsov} A.~V.,  2014, \mn@doi [\apj]
  {10.1088/0004-637X/789/1/1}, \href
  {https://ui.adsabs.harvard.edu/abs/2020arXiv200101160M/abstract} {789, 18}

\bibitem[\protect\citeauthoryear{{Diemer}, {More}  \& {Kravtsov}}{{Diemer}
  et~al.}{2013}]{Diemer2013}
{Diemer} B.,  {More} S.,   {Kravtsov} A.~V.,  2013, \mn@doi [\apj]
  {10.1088/0004-637X/766/1/25}, \href
  {https://ui.adsabs.harvard.edu/abs/2013ApJ...766...25D} {766, 25}

\bibitem[\protect\citeauthoryear{{Diemer}, {Mansfield}, {Kravtsov}  \&
  {More}}{{Diemer} et~al.}{2017}]{Diemer2017b}
{Diemer} B.,  {Mansfield} P.,  {Kravtsov} A.~V.,   {More} S.,  2017, \mn@doi
  [\apj] {10.3847/1538-4357/aa79ab}, \href
  {https://ui.adsabs.harvard.edu/abs/2017ApJ...843..140D} {843, 140}

\bibitem[\protect\citeauthoryear{{Dolag}, {Borgani}, {Murante}  \&
  {Springel}}{{Dolag} et~al.}{2009}]{Dolag2009}
{Dolag} K.,  {Borgani} S.,  {Murante} G.,   {Springel} V.,  2009, \mn@doi
  [\mnras] {10.1111/j.1365-2966.2009.15034.x}, \href
  {http://adsabs.harvard.edu/abs/2009MNRAS.399..497D} {399, 497}

\bibitem[\protect\citeauthoryear{{Donnari} et~al.,}{{Donnari}
  et~al.}{2021}]{Donnari2020}
{Donnari} M.,  et~al., 2021, \mn@doi [\mnras] {10.1093/mnras/staa3006}, \href
  {https://ui.adsabs.harvard.edu/abs/2021MNRAS.500.4004D} {500, 4004}

\bibitem[\protect\citeauthoryear{{Einasto}}{{Einasto}}{1965}]{Einasto1965}
{Einasto} J.,  1965, Trudy Astrofizicheskogo Instituta Alma-Ata, \href
  {https://ui.adsabs.harvard.edu/abs/1965TrAlm...5...87E} {5, 87}

\bibitem[\protect\citeauthoryear{{Einasto}}{{Einasto}}{1969}]{Einasto1969}
{Einasto} J.,  1969, \mn@doi [Astrophysics] {10.1007/BF01013353}, \href
  {https://ui.adsabs.harvard.edu/abs/1969Ap......5...67E} {5, 67}

\bibitem[\protect\citeauthoryear{{Fillmore} \& {Goldreich}}{{Fillmore} \&
  {Goldreich}}{1984}]{Fillmore1984}
{Fillmore} J.~A.,  {Goldreich} P.,  1984, \mn@doi [\apj] {10.1086/162070},
  \href {https://ui.adsabs.harvard.edu/abs/1984ApJ...281....1F/abstract} {281,
  1}

\bibitem[\protect\citeauthoryear{{Fong}, {Bowyer}, {Whitehead}, {Lee}, {King},
  {Applegate}  \& {McCarthy}}{{Fong} et~al.}{2018}]{Fong2018}
{Fong} M.,  {Bowyer} R.,  {Whitehead} A.,  {Lee} B.,  {King} L.,  {Applegate}
  D.,   {McCarthy} I.,  2018, \mn@doi [\mnras] {10.1093/mnras/sty1339}, \href
  {https://ui.adsabs.harvard.edu/abs/2018MNRAS.478.5366F} {478, 5366}

\bibitem[\protect\citeauthoryear{{Gao}, {Navarro}, {Cole}, {Frenk}, {White},
  {Springel}, {Jenkins}  \& {Neto}}{{Gao} et~al.}{2008}]{Gao2008}
{Gao} L.,  {Navarro} J.~F.,  {Cole} S.,  {Frenk} C.~S.,  {White} S.~D.~M.,
  {Springel} V.,  {Jenkins} A.,   {Neto} A.~F.,  2008, \mn@doi [\mnras]
  {10.1111/j.1365-2966.2008.13277.x}, \href
  {https://ui.adsabs.harvard.edu/abs/2008MNRAS.387..536G/abstract} {387, 536}

\bibitem[\protect\citeauthoryear{{Genel} et~al.,}{{Genel}
  et~al.}{2018}]{Genal2018}
{Genel} S.,  et~al., 2018, \mn@doi [\mnras] {10.1093/mnras/stx3078}, \href
  {https://ui.adsabs.harvard.edu/abs/2018MNRAS.474.3976G} {474, 3976}

\bibitem[\protect\citeauthoryear{{Graham}, {Merritt}, {Moore}, {Diemand }  \&
  {Terzi{\'c}}}{{Graham} et~al.}{2006}]{Graham2006}
{Graham} A.~W.,  {Merritt} D.,  {Moore} B.,  {Diemand } J.,   {Terzi{\'c}} B.,
  2006, \mn@doi [\aj] {10.1086/508990}, \href
  {https://ui.adsabs.harvard.edu/abs/2006AJ....132.2701G} {132, 2701}

\bibitem[\protect\citeauthoryear{{Gunn}}{{Gunn}}{1977}]{Gunn1977}
{Gunn} J.~E.,  1977, \mn@doi [\apj] {10.1086/155715}, \href
  {https://ui.adsabs.harvard.edu/abs/1977ApJ...218..592G} {218, 592}

\bibitem[\protect\citeauthoryear{{Gunn} \& {Gott}}{{Gunn} \&
  {Gott}}{1972}]{Gunn1972}
{Gunn} J.~E.,  {Gott} J.~Richard I.,  1972, \mn@doi [\apj] {10.1086/151605},
  \href {https://ui.adsabs.harvard.edu/abs/1972ApJ...176....1G} {176, 1}

\bibitem[\protect\citeauthoryear{{Huss}, {Jain}  \& {Steinmetz}}{{Huss}
  et~al.}{1999}]{Huss1999}
{Huss} A.,  {Jain} B.,   {Steinmetz} M.,  1999, \mn@doi [\apj]
  {10.1086/307161}, \href
  {https://ui.adsabs.harvard.edu/abs/1999ApJ...517...64H/abstract} {517, 64}

\bibitem[\protect\citeauthoryear{{Kaiser}}{{Kaiser}}{1986}]{Kaiser1986}
{Kaiser} N.,  1986, \mn@doi [\mnras] {10.1093/mnras/222.2.323}, \href
  {https://ui.adsabs.harvard.edu/abs/1986MNRAS.222..323K} {222, 323}

\bibitem[\protect\citeauthoryear{{Klein} et~al.,}{{Klein}
  et~al.}{2019}]{Klein2019}
{Klein} M.,  et~al., 2019, \mn@doi [\mnras] {10.1093/mnras/stz1463}, \href
  {https://ui.adsabs.harvard.edu/abs/2019MNRAS.488..739K} {488, 739}

\bibitem[\protect\citeauthoryear{{Koester} et~al.,}{{Koester}
  et~al.}{2007}]{Koester2007}
{Koester} B.~P.,  et~al., 2007, \mn@doi [\apj] {10.1086/509599}, \href
  {https://ui.adsabs.harvard.edu/abs/2007ApJ...660..239K} {660, 239}

\bibitem[\protect\citeauthoryear{{Kravtsov} \& {Borgani}}{{Kravtsov} \&
  {Borgani}}{2012}]{Kravtsov2012}
{Kravtsov} A.~V.,  {Borgani} S.,  2012, \mn@doi [\araa]
  {10.1146/annurev-astro-081811-125502}, \href
  {https://ui.adsabs.harvard.edu/abs/2012ARA&A..50..353K} {50, 353}

\bibitem[\protect\citeauthoryear{{Lau}, {Herter}, {Morris}, {Li}  \&
  {Adams}}{{Lau} et~al.}{2015}]{Lau2015}
{Lau} R.~M.,  {Herter} T.~L.,  {Morris} M.~R.,  {Li} Z.,   {Adams} J.~D.,
  2015, \mn@doi [Science] {10.1126/science.aaa2208}, \href
  {http://adsabs.harvard.edu/abs/2015Sci...348..413L} {348, 413}

\bibitem[\protect\citeauthoryear{{Ludlow}, {Navarro}, {White},
  {Boylan-Kolchin}, {Springel}, {Jenkins}  \& {Frenk}}{{Ludlow}
  et~al.}{2011}]{Ludlow2011}
{Ludlow} A.~D.,  {Navarro} J.~F.,  {White} S. D.~M.,  {Boylan-Kolchin} M.,
  {Springel} V.,  {Jenkins} A.,   {Frenk} C.~S.,  2011, \mn@doi [\mnras]
  {10.1111/j.1365-2966.2011.19008.x}, \href
  {https://ui.adsabs.harvard.edu/abs/2011MNRAS.415.3895L} {415, 3895}

\bibitem[\protect\citeauthoryear{{Mansfield}, {Kravtsov}  \&
  {Diemer}}{{Mansfield} et~al.}{2017}]{Mansfield2017}
{Mansfield} P.,  {Kravtsov} A.~V.,   {Diemer} B.,  2017, \mn@doi [\apj]
  {10.3847/1538-4357/aa7047}, \href
  {https://ui.adsabs.harvard.edu/abs/2017ApJ...841...34M/abstract} {841, 21}

\bibitem[\protect\citeauthoryear{{Marinacci} et~al.,}{{Marinacci}
  et~al.}{2018}]{Marinacci2018}
{Marinacci} F.,  et~al., 2018, \mnras, \href
  {https://ui.adsabs.harvard.edu/abs/2018MNRAS.480.5113M/abstract} {480, 5113}

\bibitem[\protect\citeauthoryear{{Markevitch} \& {Vikhlinin}}{{Markevitch} \&
  {Vikhlinin}}{2007}]{Markevitch2007}
{Markevitch} M.,  {Vikhlinin} A.,  2007, \mn@doi [\physrep]
  {10.1016/j.physrep.2007.01.001}, \href
  {https://ui.adsabs.harvard.edu/abs/2007PhR...443....1M} {443, 1}

\bibitem[\protect\citeauthoryear{{McCarthy}, {Schaye}, {Bird}  \& {Le
  Brun}}{{McCarthy} et~al.}{2017}]{McCarthy2017}
{McCarthy} I.~G.,  {Schaye} J.,  {Bird} S.,   {Le Brun} A. M.~C.,  2017,
  \mn@doi [\mnras] {10.1093/mnras/stw2792}, \href
  {https://ui.adsabs.harvard.edu/abs/2017MNRAS.465.2936M} {465, 2936}

\bibitem[\protect\citeauthoryear{{Merritt}, {Graham}, {Moore}, {Diemand }  \&
  {Terzi{\'c}}}{{Merritt} et~al.}{2006}]{Merritt2006}
{Merritt} D.,  {Graham} A.~W.,  {Moore} B.,  {Diemand } J.,   {Terzi{\'c}} B.,
  2006, \mn@doi [\aj] {10.1086/508988}, \href
  {https://ui.adsabs.harvard.edu/abs/2006AJ....132.2685M} {132, 2685}

\bibitem[\protect\citeauthoryear{{More}, {Diemer}  \& {Kravtsov}}{{More}
  et~al.}{2015}]{More2015}
{More} S.,  {Diemer} B.,   {Kravtsov} A.~V.,  2015, \mn@doi [\apj]
  {10.1088/0004-637X/810/1/36}, \href
  {https://ui.adsabs.harvard.edu/abs/2015ApJ...810...36M} {810, 36}

\bibitem[\protect\citeauthoryear{{More} S.~{Miyatake} et~al.,}{{More}
  et~al.}{2016}]{More2016}
{More} S.~{Miyatake} H.,  et~al., 2016, \mn@doi [\apj]
  {10.3847/0004-637X/825/1/39}, \href
  {https://ui.adsabs.harvard.edu/abs/2016ApJ...825...39M/abstract} {825, 18}

\bibitem[\protect\citeauthoryear{{Murata}, {Sunayama}, {Oguri}, {More},
  {Nishizawa}, {Nishimichi}  \& {Osato}}{{Murata} et~al.}{2020}]{Murata2020}
{Murata} R.,  {Sunayama} T.,  {Oguri} M.,  {More} S.,  {Nishizawa} A.~J.,
  {Nishimichi} T.,   {Osato} K.,  2020, \mn@doi [\pasj] {10.1093/pasj/psaa041},
  \href {https://ui.adsabs.harvard.edu/abs/2020PASJ...72...64M} {72, 64}

\bibitem[\protect\citeauthoryear{{Naiman} et~al.,}{{Naiman}
  et~al.}{2018}]{Naiman2018}
{Naiman} J.~P.,  et~al., 2018, \mn@doi [\mnras] {10.1093/mnras/sty618}, \href
  {https://ui.adsabs.harvard.edu/abs/2018MNRAS.477.1206N/abstraPillct} {477,
  1206}

\bibitem[\protect\citeauthoryear{{Navarro}, {Frenk}  \& {White}}{{Navarro}
  et~al.}{1996}]{Navarro1996}
{Navarro} J.~F.,  {Frenk} C.~S.,   {White} S. D.~M.,  1996, \mn@doi [\apj]
  {10.1086/177173}, \href
  {https://ui.adsabs.harvard.edu/abs/1996ApJ...462..563N} {462, 563}

\bibitem[\protect\citeauthoryear{{Navarro} et~al.,}{{Navarro}
  et~al.}{2010}]{Navarro2010}
{Navarro} J.~F.,  et~al., 2010, \mn@doi [\mnras]
  {10.1111/j.1365-2966.2009.15878.x}, \href
  {https://ui.adsabs.harvard.edu/abs/2010MNRAS.402...21N} {402, 21}

\bibitem[\protect\citeauthoryear{{Nelson} et~al.,}{{Nelson}
  et~al.}{2018}]{Nelson2018}
{Nelson} D.,  et~al., 2018, \mnras, \href
  {https://ui.adsabs.harvard.edu/abs/2018MNRAS.475..624N/abstract} {475, 624}

\bibitem[\protect\citeauthoryear{{Pakmor}, {Volker}, {Bauer}, {Mocz}, {Munoz},
  {Ohlmann}, {Schaal}  \& {Zhu}}{{Pakmor} et~al.}{2016}]{Pakmor2016}
{Pakmor} R.,  {Volker} S.,  {Bauer} A.,  {Mocz} P.,  {Munoz} D.~J.,  {Ohlmann}
  S.~T.,  {Schaal} K.,   {Zhu} C.,  2016, \mn@doi [\mnras]
  {10.1093/mnras/stv2380}, \href
  {https://ui.adsabs.harvard.edu/abs/2016MNRAS.455.1134P/abstract} {455, 1134}

\bibitem[\protect\citeauthoryear{{Peebles}}{{Peebles}}{1980}]{Peebles1980}
{Peebles} P.~J.~E.,  1980, {The large-scale structure of the universe}.
Princeton University Press

\bibitem[\protect\citeauthoryear{{Pike}, {Kay}, {Newton}, {Thomas}  \&
  {Jenkins}}{{Pike} et~al.}{2014}]{Pike2014}
{Pike} S.~R.,  {Kay} S.~T.,  {Newton} R. D.~A.,  {Thomas} P.~A.,   {Jenkins}
  A.,  2014, \mn@doi [\mnras] {10.1093/mnras/stu1788}, \href
  {https://ui.adsabs.harvard.edu/abs/2014MNRAS.445.1774P} {445, 1774}

\bibitem[\protect\citeauthoryear{{Pillepich} et~al.,}{{Pillepich}
  et~al.}{2018a}]{Pillepich2018a}
{Pillepich} A.,  et~al., 2018a, \mn@doi [\mnras] {10.1093/mnras/stx2656}, \href
  {http://adsabs.harvard.edu/abs/2018MNRAS.473.4077P} {473, 4077}

\bibitem[\protect\citeauthoryear{{Pillepich} et~al.,}{{Pillepich}
  et~al.}{2018b}]{Pillepich2018b}
{Pillepich} A.,  et~al., 2018b, \mn@doi [\mnras] {10.1093/mnras/stx3112}, \href
  {https://ui.adsabs.harvard.edu/abs/2018MNRAS.475..648P/abstract} {475, 648}

\bibitem[\protect\citeauthoryear{{Planck Collaboration} et~al.,}{{Planck
  Collaboration} et~al.}{2016}]{PlanckCollaborationXIII2016}
{Planck Collaboration} et~al., 2016, \mn@doi [\aap]
  {10.1051/0004-6361/201525830}, \href
  {https://ui.adsabs.harvard.edu/abs/2016A%26A...594A..13P/abstract} {594, 63}

\bibitem[\protect\citeauthoryear{{Rees}}{{Rees}}{1977}]{Rees1977}
{Rees} M.~J.,  1977, in {Tinsley} B.~M.,  {Larson} Richard B.~Gehret D.~C.,
  eds, Evolution of Galaxies and Stellar Populations. p.~339

\bibitem[\protect\citeauthoryear{{Reiprich}, {Basu}, {Ettori}, {Israel},
  {Lovisari}, {Molendi}, {Pointecouteau}  \& {Roncarelli}}{{Reiprich}
  et~al.}{2013}]{Reiprich2013}
{Reiprich} T.~H.,  {Basu} K.,  {Ettori} S.,  {Israel} H.,  {Lovisari} L.,
  {Molendi} S.,  {Pointecouteau} E.,   {Roncarelli} M.,  2013, \mn@doi [\ssr]
  {10.1007/s11214-013-9983-8}, \href
  {https://ui.adsabs.harvard.edu/abs/2013SSRv..177..195R} {177, 195}

\bibitem[\protect\citeauthoryear{{Rodriguez-Gomez} et~al.,}{{Rodriguez-Gomez}
  et~al.}{2015}]{Rodriguez-Gomez2015}
{Rodriguez-Gomez} V.,  et~al., 2015, \mn@doi [\mnras] {10.1093/mnras/stv264},
  \href {https://ui.adsabs.harvard.edu/abs/2015MNRAS.449...49R/abstract} {449,
  49}

\bibitem[\protect\citeauthoryear{{Rozo}, {Wechsler}, {Koester}, {Evrard}  \&
  {McKay}}{{Rozo} et~al.}{2007}]{Rozo2007}
{Rozo} E.,  {Wechsler} R.~H.,  {Koester} B.~P.,  {Evrard} A.~E.,   {McKay}
  T.~A.,  2007, arXiv e-prints, \href
  {https://ui.adsabs.harvard.edu/abs/2007astro.ph..3574R} {pp
  astro--ph/0703574}

\bibitem[\protect\citeauthoryear{{Rykoff} et~al.,}{{Rykoff}
  et~al.}{2014}]{Rykoff2014}
{Rykoff} E.~S.,  et~al., 2014, \mn@doi [\apj] {10.1088/0004-637X/785/2/104},
  \href {https://ui.adsabs.harvard.edu/abs/2014ApJ...785..104R} {785, 104}

\bibitem[\protect\citeauthoryear{{Schaye} et~al.,}{{Schaye}
  et~al.}{2015}]{Schaye2015}
{Schaye} J.,  et~al., 2015, \mn@doi [\mnras] {10.1093/mnras/stu2058}, \href
  {http://adsabs.harvard.edu/abs/2015MNRAS.446..521S} {446, 521}

\bibitem[\protect\citeauthoryear{{Sheth}, {Mo}  \& {Tormen}}{{Sheth}
  et~al.}{2001}]{Sheth2001}
{Sheth} R.~K.,  {Mo} H.~J.,   {Tormen} G.,  2001, \mn@doi [\mnras]
  {10.1046/j.1365-8711.2001.04006.x}, \href
  {https://ui.adsabs.harvard.edu/abs/2001MNRAS.323....1S} {323, 1}

\bibitem[\protect\citeauthoryear{{Shi}}{{Shi}}{2016}]{Shi2016b}
{Shi} X.,  2016, \mn@doi [\mnras] {10.1093/mnras/stw1418}, \href
  {https://ui.adsabs.harvard.edu/abs/2016MNRAS.461.1804S/abstract} {461, 1804}

\bibitem[\protect\citeauthoryear{{Shin} et~al.,}{{Shin}
  et~al.}{2019}]{Shin2019}
{Shin} T.,  et~al., 2019, \mn@doi [\mnras] {10.1093/mnras/stz1434}, \href
  {https://ui.adsabs.harvard.edu/abs/2019MNRAS.487.2900S/abstract} {487, 2900}

\bibitem[\protect\citeauthoryear{{Springel}}{{Springel}}{2010}]{Springel2010}
{Springel} V.,  2010, \mn@doi [\mnras] {10.1111/j.1365-2966.2009.15715.x},
  \href {http://adsabs.harvard.edu/abs/2010MNRAS.401..791S} {401, 791}

\bibitem[\protect\citeauthoryear{{Springel} \& {Hernquist}}{{Springel} \&
  {Hernquist}}{2003}]{Springel2003}
{Springel} V.,  {Hernquist} L.,  2003, \mn@doi [\mnras]
  {10.1046/j.1365-8711.2003.06206.x}, \href
  {http://adsabs.harvard.edu/abs/2003MNRAS.339..289S} {339, 289}

\bibitem[\protect\citeauthoryear{{Springel}, {White}, {Tormen}  \&
  {Kauffmann}}{{Springel} et~al.}{2001}]{Springel2001}
{Springel} V.,  {White} S.~D.~M.,  {Tormen} G.,   {Kauffmann} G.,  2001,
  \mn@doi [\mnras] {10.1046/j.1365-8711.2001.04912.x}, \href
  {http://adsabs.harvard.edu/abs/2001MNRAS.328..726S} {328, 726}

\bibitem[\protect\citeauthoryear{{Springel} et~al.,}{{Springel}
  et~al.}{2018}]{Springel2018}
{Springel} V.,  et~al., 2018, \mn@doi [\mnras] {10.1093/mnras/stx3304}, \href
  {https://ui.adsabs.harvard.edu/abs/2018MNRAS.475..676S/abstract} {475, 676}

\bibitem[\protect\citeauthoryear{{Stadel}, {Potter}, {Moore}, {Diemand},
  {Madau}, {Zemp}, {Kuhlen}  \& {Quilis}}{{Stadel} et~al.}{2009}]{Stadel2009}
{Stadel} J.,  {Potter} D.,  {Moore} B.,  {Diemand} J.,  {Madau} P.,  {Zemp} M.,
   {Kuhlen} M.,   {Quilis} V.,  2009, \mn@doi [\mnras]
  {10.1111/j.1745-3933.2009.00699.x}, \href
  {https://ui.adsabs.harvard.edu/abs/2009MNRAS.398L..21S} {398, L21}

\bibitem[\protect\citeauthoryear{{Torrey}, {Vogelsberger}, {Genel}, {Sijacki},
  {Springel}  \& {Hernquist}}{{Torrey} et~al.}{2014}]{Torrey2014}
{Torrey} P.,  {Vogelsberger} M.,  {Genel} S.,  {Sijacki} D.,  {Springel} V.,
  {Hernquist} L.,  2014, \mn@doi [\mnras] {10.1093/mnras/stt2295}, \href
  {http://adsabs.harvard.edu/abs/2014MNRAS.438.1985T} {438, 1985}

\bibitem[\protect\citeauthoryear{{Vikhlinin}, {Kravtsov}, {Forman}, {Jones},
  {Markevitch}, {Murray}  \& {Van Speybroeck}}{{Vikhlinin}
  et~al.}{2006}]{Vikhlinin2006}
{Vikhlinin} A.,  {Kravtsov} A.,  {Forman} W.,  {Jones} C.,  {Markevitch} M.,
  {Murray} S.~S.,   {Van Speybroeck} L.,  2006, \mn@doi [\apj]
  {10.1086/500288}, \href
  {https://ui.adsabs.harvard.edu/abs/2006ApJ...640..691V/abstract} {640, 691}

\bibitem[\protect\citeauthoryear{{Vogelsberger}, {White}, {Mohayaee}  \&
  {Springel}}{{Vogelsberger} et~al.}{2009}]{Vogelsberger2009}
{Vogelsberger} M.,  {White} S. D.~M.,  {Mohayaee} R.,   {Springel} V.,  2009,
  \mn@doi [\mnras] {10.1111/j.1365-2966.2009.15615.x}, \href
  {https://ui.adsabs.harvard.edu/abs/2009MNRAS.400.2174V} {400, 2174}

\bibitem[\protect\citeauthoryear{{Vogelsberger}, {Genel}, {Sijacki}, {Torrey},
  {Springel}  \& {Hernquist}}{{Vogelsberger} et~al.}{2013}]{Vogelsberger2013}
{Vogelsberger} M.,  {Genel} S.,  {Sijacki} D.,  {Torrey} P.,  {Springel} V.,
  {Hernquist} L.,  2013, \mn@doi [\mnras] {10.1093/mnras/stt1789}, \href
  {http://adsabs.harvard.edu/abs/2013MNRAS.436.3031V} {436, 3031}

\bibitem[\protect\citeauthoryear{{Vogelsberger} et~al.,}{{Vogelsberger}
  et~al.}{2014a}]{Vogelsberger2014b}
{Vogelsberger} M.,  et~al., 2014a, \mn@doi [\mnras] {10.1093/mnras/stu1536},
  \href {http://adsabs.harvard.edu/abs/2014MNRAS.444.1518V} {444, 1518}

\bibitem[\protect\citeauthoryear{{Vogelsberger} et~al.,}{{Vogelsberger}
  et~al.}{2014b}]{Vogelsberger2014a}
{Vogelsberger} M.,  et~al., 2014b, \mn@doi [\nat] {10.1038/nature13316}, \href
  {http://adsabs.harvard.edu/abs/2014Natur.509..177V} {509, 177}

\bibitem[\protect\citeauthoryear{{Vogelsberger} et~al.,}{{Vogelsberger}
  et~al.}{2018}]{Vogelsberger2018}
{Vogelsberger} M.,  et~al., 2018, \mn@doi [\mnras] {10.1093/mnras/stx2955},
  \href {http://adsabs.harvard.edu/abs/2018MNRAS.474.2073V} {474, 2073}

\bibitem[\protect\citeauthoryear{{Vogelsberger}, {Marinacci}, {Torrey}  \&
  {Puchwein}}{{Vogelsberger} et~al.}{2020}]{Vogelsberger2020}
{Vogelsberger} M.,  {Marinacci} F.,  {Torrey} P.,   {Puchwein} E.,  2020,
  \mn@doi [Nature Reviews Physics] {10.1038/s42254-019-0127-2}, \href
  {https://ui.adsabs.harvard.edu/abs/2020NatRP...2...42V} {2, 42}

\bibitem[\protect\citeauthoryear{{Weinberger} et~al.,}{{Weinberger}
  et~al.}{2017}]{Weinberger2017}
{Weinberger} R.,  et~al., 2017, \mn@doi [\mnras] {10.1093/mnras/stw2944}, \href
  {https://ui.adsabs.harvard.edu/abs/2017MNRAS.465.3291W/abstract} {465, 3291}

\bibitem[\protect\citeauthoryear{{Weinberger}, {Springel}  \&
  {Pakmor}}{{Weinberger} et~al.}{2020}]{Weinberger2020}
{Weinberger} R.,  {Springel} V.,   {Pakmor} R.,  2020, \mn@doi [The
  Astrophysical Journal Supplement Series] {10.3847/1538-4365/ab908c}, \href
  {https://ui.adsabs.harvard.edu/abs/2020ApJS..248...32W/abstract} {248, 39}

\bibitem[\protect\citeauthoryear{{White} \& {Rees}}{{White} \&
  {Rees}}{1978}]{White1978}
{White} S.~D.~M.,  {Rees} M.~J.,  1978, \mn@doi [\mnras]
  {10.1093/mnras/183.3.341}, \href
  {https://ui.adsabs.harvard.edu/abs/1978MNRAS.183..341W} {183, 341}

\bibitem[\protect\citeauthoryear{{Xhakaj}, {Diemer}, {Leauthaud}, {Wasserman},
  {Huang}, {Luo}, {Adhikari}  \& {Singh}}{{Xhakaj} et~al.}{2020}]{Xhakaj2020}
{Xhakaj} E.,  {Diemer} B.,  {Leauthaud} A.,  {Wasserman} A.,  {Huang} S.,
  {Luo} Y.,  {Adhikari} S.,   {Singh} S.,  2020, \mn@doi [\mnras]
  {10.1093/mnras/staa3046}, \href
  {https://ui.adsabs.harvard.edu/abs/2020MNRAS.499.3534X} {499, 3534}

\bibitem[\protect\citeauthoryear{{Zhang}, {Zhuravleva}, {Kravtsov}  \&
  {Churazov}}{{Zhang} et~al.}{2021}]{Zhang2021}
{Zhang} C.,  {Zhuravleva} I.,  {Kravtsov} A.,   {Churazov} E.,  2021, arXiv
  e-prints, \href {https://ui.adsabs.harvard.edu/abs/2021arXiv210303850Z} {p.
  arXiv:2103.03850}

\bibitem[\protect\citeauthoryear{{Z{\"u}rcher} \& {More}}{{Z{\"u}rcher} \&
  {More}}{2019}]{Zuercher2019}
{Z{\"u}rcher} D.,  {More} S.,  2019, \mn@doi [\apj] {10.3847/1538-4357/ab08e8},
  \href {https://ui.adsabs.harvard.edu/abs/2019ApJ...874..184Z} {874, 184}

\makeatother
\end{thebibliography}

\appendix
\section{Numerical convergence}
\label{apx:convergence}
Here we examine the effects of resolution on our results to ensure that they are robust.  TNG300 and TNG300-DM were run at three resolution levels with $2500^3$, $1250^3$ and $625^3$ dark matter particles.  To compare the resolution levels, we compute $R_{\rm sp}$ for the dark matter profiles in TNG300-1, TNG300-2 and TNG300-3.

The intermediate resolution, TNG300-2, reduces the number of resolution elements by a factor $8$ and the spatial resolution by a factor $2$, and the lowest resolution run, TNG300-3, reduces the resolution elements and spatial resolution by a further factor of $8$ and $2$, respectively.

\begin{table*}
\begin{tabular}{c|c|cccc|cccccccc}
     Simulation & $z$ & \multicolumn{4}{c|}{$\log_{10}\left(M_{\rm halo}\,/\,{{\rm M}_{\odot}}\right)$} & \multicolumn{8}{c}{$\Gamma$} \\
     & & \!{13.0-13.5}\! & \!{13.5-14.0}\! &  \!{14.0-14.5}\! & \!{14.5-15.0}\! & \!0.0-0.5\! & \!0.5-1.0\! & \!1.0-1.5\! & \!1.5-2.0\! & \!2.0-2.5\! & \!2.5-3.0\! & \!3.0-4.0\! & \!4.0-5.0\! \\
     \hline
     &  $0.0$  & 770 & 394 & 185 & 47 & 
                  109 & 361 & 316 & 237 & 154 & 90  & 87  & 25 \\
     TNG300-1 &  $0.5$  & 991 & 444 & 119 & 13 & 
                  83  & 197 & 267 & 222 & 182 & 176 & 228  & 127 \\
     &  $1.0$  & 1005 & 336 & 50  & 2  & 
                  39   & 102 & 152 & 145 & 168 & 157 & 243 & 198 \\
    \hline
                & $0.0$  & 782 & 391 & 186 & 47 
                & 115 & 331 & 356 & 217 & 156 & 99 & 92 & 23 \\
       TNG300-2 & $0.5$  & 1001 & 446 & 117 & 13
                & 83   & 195 & 261 & 220 & 177 & 166 & 251 & 143 \\
                & $1.0$  & 1008 & 329 & 48  & 2 
                & 38   & 105 & 142 & 140 & 162 & 149 & 265  & 169 \\
      \hline
                & $0.0$  & 787  & 417 & 177 & 44
                & 106  & 358 & 347 & 222 & 154 & 96  & 101  & 29 \\
       TNG300-3 & $0.5$  & 978 & 456 & 117 & 11 
                & 69  & 184 & 281 & 237 & 184 & 172 & 214 & 126  \\
                & $1.0$  & 1026 & 334 & 43  & 2 
                & 29   & 103 & 140 & 199 & 155 & 146 & 267 &  165\\
\end{tabular}
\caption{The number of haloes found in the high, medium and low resolution (levels 1, 2 and 3) hydrodynamic (TNG300) simulations for each mass and accretion rate threshold for $z=0, 0.5$ and $1$.  This is the number of friends-of-friends groups in the given mass or accretion range, and haloes within $10\times R_{\rm200m}$ of a larger halo are not counted.}
\label{table:res_sample_size}
\end{table*}

\begin{figure*}
    \centering
    \includegraphics[width=\linewidth]{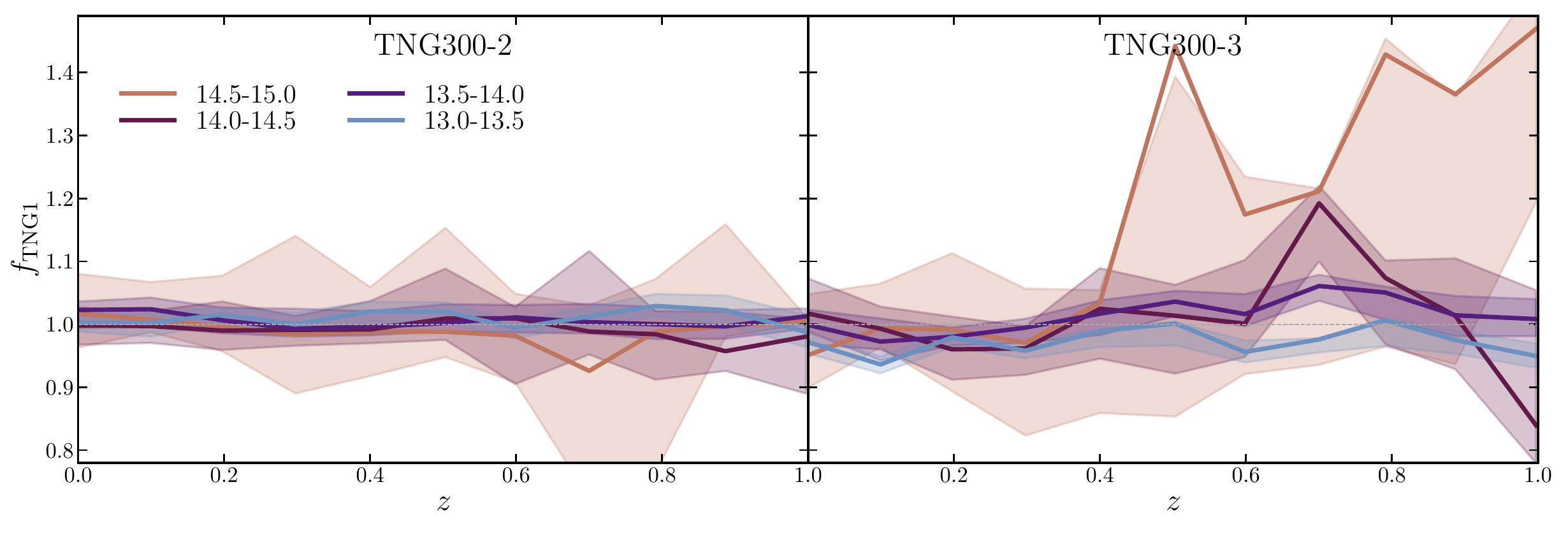}
    \vspace{-.5cm}
    \caption{Here we compare results of the dark matter mass profiles between the resolution levels of TNG300.  Each resolution level has $\left(\frac{1}{2}\right)^3$ as many particles as the previous resolution level.  We stack the density profiles for haloes with $\log_{10}\left(M_{200\rm{m}}/{\rm M}_{\odot}\right)$ in ${13}-{13.5}$, ${13.5}-{14}$, ${14}-{14.5}$ and ${14.5}-{15}$ and compute the splashback radius of the median profile, shown by the solid lines.  To estimate error, we use a bootstrap method and show the 16th and 84th percentiles as the shaded band around each line.  We show the fractional difference in $R_{\rm sp}$ between TNG300-1 and TNG300-2 (left) or TNG300-3 (right) and do not find a significant difference.  TNG300-1 deviates at higher redshift for high-mass haloes, but this is well-behaved in TNG300-2.  $R_{\rm sp}$ also pushes against (or past) the bootstrap error bands in the high mass bin for TNG300-3 where the sample size is small.}
    \label{fig:redshift_evolution_convergence}
\end{figure*}

Figure \ref{fig:redshift_evolution_convergence} shows the fractional difference of $R_{\rm sp}$ between the lower resolution runs, TNG300-2 and TNG300-3, and the high resolution run, TNG300-1, in each mass bin as a function of redshift for each resolution level as described in Section \ref{sec:results_redshift}.  TNG300-3 deviates from the expected behaviour at high redshift for the most massive haloes, although the error here is also large.  This is likely due to the small number of haloes in this range combined with the lower resolution.  In general, however, all three simulations yield results in good agreement.

\begin{figure*}
    \centering
    \includegraphics[width=\linewidth]{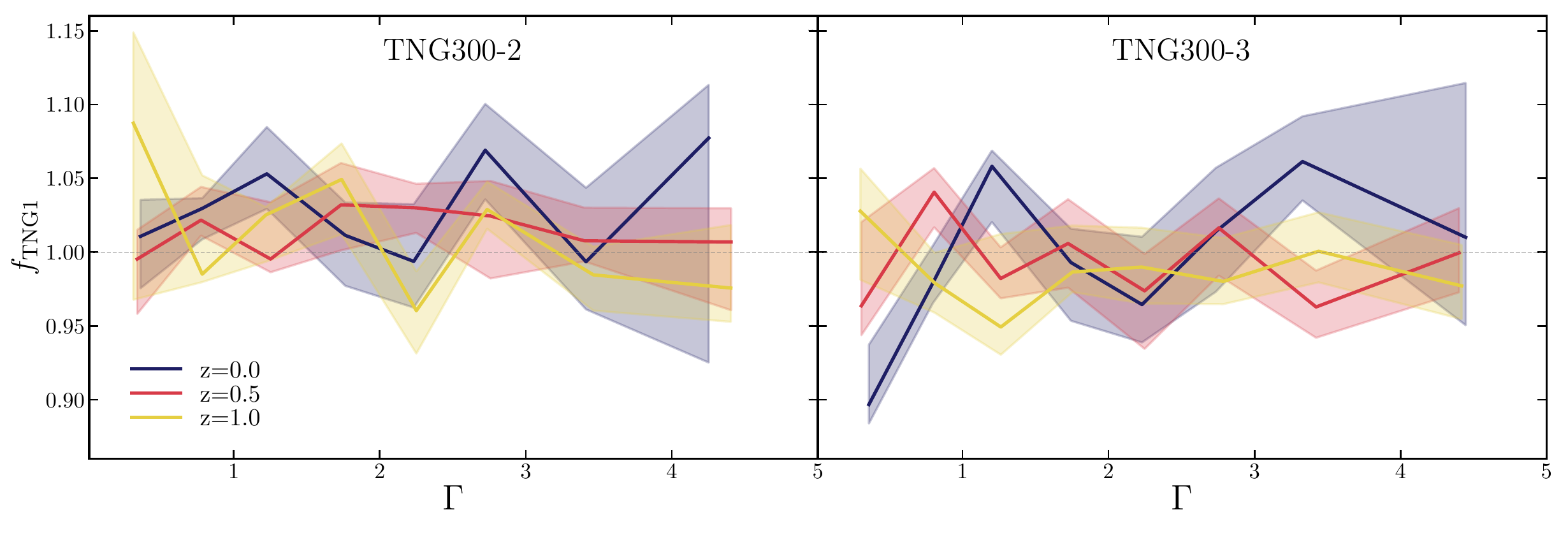}
    \vspace{-.5cm}
    \caption{We show the fractional difference between $R_{\rm sp}$ in TNG300-1 and TNG300-2 (left) or TNG300-3 (right) calculated as a function of accretion rate $\Gamma$.  The colours show several redshifts.  Each resolution level has $\left(\frac{1}{2}\right)^3$ as many particles as the previous resolution level.  The haloes are stacked based on accretion rates in eight bins between $0$ and $5$.  We compute the splashback radius of the median profile for each accretion range as described in Section \ref{sec:results_accretion}.}
    \label{fig:gamma_convergence}
\end{figure*}

Figure \ref{fig:gamma_convergence} shows the fractional difference of $R_{\rm sp}$ between resolution levels as a function of accretion rate for redshifts between $0$ and $1$.  We do not find any significant deviation in the results in the three resolution levels.  This indicates that resolution does not play a role in our main results.

\label{lastpage}

\end{document}